\documentclass[12pt,preprint]{aastex}
\usepackage{graphicx}
\newcommand{\De}{\ensuremath{{\rm D}}}
\newcommand{\D}{\ensuremath{\null^2{\rm H}}}
\newcommand{\Tr}{\ensuremath{{\rm T}}}
\newcommand{\T}{\ensuremath{\null^3{\rm H}}}
\newcommand{\Het}{\ensuremath{\null^3{\rm He}}}
\newcommand{\He}{\ensuremath{\null^4{\rm He}}}
\newcommand{\Be}{\ensuremath{\null^7{\rm Be}}}
\newcommand{\Lis}{\ensuremath{\null^6{\rm Li}}}
\newcommand{\Li}{\ensuremath{\null^7{\rm Li}}}

\begin{document}

\title{{\bf Primordial Nucleosynthesis with varying fundamental
    constants: A semi-analytical approach}} 

\author{Susana J. Landau \altaffilmark{1,2}} 
\email{slandau@df.uba.ar} 
\and 
\author{Mercedes E. Mosquera \altaffilmark{1}} 
\email{mmosquera@carina.fcaglp.unlp.edu.ar} 
\and  
\author{Hector Vucetich \altaffilmark{1}}
\email{vucetich@fcaglp.unlp.edu.ar}

\altaffiltext{1}{Facultad de Ciencias Astron\'{o}micas y
  Geof\'{\i}sicas. Universidad Nacional de La Plata. Paseo del 
Bosque S/N 1900 La Plata, Argentina}
\altaffiltext{2}{Departamento de F{\'\i}sica, FCEyN, Universidad de
  Buenos Aires, Ciudad Universitaria - Pab. 1, 1428 Buenos Aires,
  Argentina} 

\keywords{primordial nucleosynthesis, varying fundamental constants, cosmology}

\begin{abstract}
Using the semi-analytic method proposed by \citet{Esma91} we calculate
the abundances of the light elements produced during primordial
nucleosynthesis assuming that the gauge coupling constants of the
fundamental interactions may vary.  We analyze the dependence of the
nucleon masses, nuclear binding energies and cross sections involved
in the calculation of the abundances with the fundamental constants
assuming the chiral limit of QCD. The abundances of light elements as
a function of the fundamental constants are obtained. Finally, using
the observational data of $\De$, $\Het$, $\He$ and $\Li$ we estimate
constraints on the variation of the fundamental constants between the
primordial nucleosynthesis and the present. All observational
abundances and the WMAP estimate of the baryon density, can be fitted
to the theoretical predictions with varying coupling constants. The
possible systematic errors in the observational data, precludes from
stronger conclusions.
\end{abstract}

\section{Introduction}
\label{sec:Intro}

Big Bang Nucleosynthesis (BBN) is one of the most important tools to
study the early universe. The model is simple and has only one free
parameter, the density of baryonic matter, which can be determined by
comparison between theoretical calculations and observations of the
abundances of the light elements.  On the other hand, data on cosmic
microwave background (CMB) provide an alternative, independent method
for determining $\Omega_B h^2$ \citep{wmapest}. Recently, the
concordance between both methods has been investigated by many authors
\citep{cyburt03,romano03, cuoco03,cyburt04,coc04,Vangioni04}. From the
WMAP baryon density , the predicted abundances are highly consistent
with the observed $\De$ but not with $\He$ and $\Li$. They are
produced more than observed. Such discrepancy is usually ascribed to
non reported systematic errors in the observations of $\He$ and
$\Li$. Indeed, more realistic determinations of the $\He$ uncertainty
implies a baryon density in line with the WMAP estimate
\citep{cyburt04,Olive04}. On the other hand, \citet{richard05} have
pointed out that a better understanding of turbulent transport in the
radiative zones of the stars is needed for a better determination of
the $\Li$ abundance. However, if the systematic errors of $\He$ and
$\Li$ are correctly estimated, we may have insight into new physics
beyond the minimal BBN model, for example: new neutron lifetime
\citep{Mathews05}, super WIMP scenario \citep{Feng03}, lepton asymmetry
\citep{ichi04b} and varying constants
\citep{Iguri99,Nollet,Ichi02,ichi04}. Therefore, BBN is not only one
of the most important tests of the Big Bang theory, but it is also
useful to obtain stringent constraints on deviations from standard
cosmology and on alternative theories to the Standard Model of
fundamental interactions (SM).

Among these theories, there are some in which the gauge coupling
constants may vary over cosmological time scales like string derived
field theories \citep{Wu86,Maeda88,Barr88,DP94,DPV2002a,DPV2002b},
related brane-world theories
\citep{Youm2001a,Youm2001b,branes03a,branes03b}, {and} (related or
not) Kaluza-Klein theories
\citep{Kaluza,Klein,Weinberg83,GT85,OW97}. On the other hand, a
theoretical framework in which only the fine structure constant varies
was developed by \citet{Bekenstein82} and improved by
\citet{BSM02}. This model was generalized in order to study the time
variation of the strong coupling constant \citep{CLV01}. Different
versions of the theories mentioned above predict different time
behaviors of the gauge coupling constants. Thus, bounds obtained from
astronomical and geophysical data are an important tool to test the
validity of these theories.

The experimental research can be grouped into astronomical and local
methods. The latter ones include geophysical methods such as the
natural nuclear reactor that operated about $1.8\ 10^9$ years ago in
Oklo, Gabon \citep{DD96,Fujii00,Fujii02}, the analysis of natural
long-lived $\beta$ decayers in geological minerals and meteorites
\citep{Dyson67,SV90,Smolliar96} and laboratory measurements such as
comparisons of rates between clocks with different atomic number
\citep{PTM95,Sortais00,Marion03}. The astronomical methods are based
mainly in the analysis of spectra form high-redshift quasar absorption
systems
\citep{CS95,VPI96,Webb99,Webb01,Murphy01a,Murphy01b,LE02,Ivanchik02,%
Murphy03b,Ivanchik03,Bahcall04}.  Although, most of the previous
mentioned experimental data gave null results, evidence of time
variation of the fine structure constant was reported recently from
high-redshift quasar absorption systems
\citep{Webb99,Webb01,Murphy01a,Murphy01b,Murphy03b,Ivanchik03}. However,
other recent independent analysis of similar data
\citep{MVB04,QRL04,Bahcall04,Srianand04} found no variation. On the
other hand, measurements of molecular hydrogen \citep{Ivanchik02,
Ivanchik03} reported a variation of the proton to electron mass $\mu =
\frac{m_p}{m_e}$
 
The time variation of the gauge coupling constants in the early
universe can be constrained using data from the Cosmic Microwave
Background (CMB) \citep{BCW01,AV00,Martins02,Rocha03} and the
primordial abundances of light elements  \citep{Iguri99,Nollet,Ichi02,ichi04}.


The prediction of the light elements abundances (\He, $\De$, \Li)
produced during the first minutes of the universe can be calculated
using numerical \citep{Wagoner73,Kawano92} and analytical
\citep{Esma91,Mukhanov03} methods. \citet{Ichi02} modified the public
code in order to analyze the BBN scenario with varying gauge coupling
constants. They considered a theoretical model taken from string theory
where the variation of the coupling constant is related to the
expectation values of the dilaton field and compared with
observational data.  In consequence, the results they obtained are
restricted to the validity of this model. Furthermore, numerical
calculations of the theoretical abundances of the light elements
allowing only a variation of the fine structure constant were
performed by different authors \citep{Iguri99,Nollet,ichi04}. On the
other hand, an analytical study of $\He$ abundance including variation
of the gauge coupling constants was performed by \citet{mueller04}. Moreover, the change in the abundance of $\He$ due to variable mass in 5 dimensional theories was analyzed by \citet{ATV96}. Finally, the effect of considering non extensive thermostatistics has been analyzed by various authors \citep{TVP97,PTV01,PT01}.

In this work, we follow the semi-analytical method proposed by
\citet{Esma91} to study the effect of a possible variation of the
values of the three gauge coupling constants of the Standard Model of
Particles Interactions (SM)
between primordial nucleosynthesis and the present. Even though, the
semi-analytical method gives results one order of magnitude less
accurate that the calculations performed with the numerical code, it
is very useful to find out the dependence of the abundances and
temperatures with the fundamental constants, which is one of the
principal aims of this work.

We will not assume any of the theoretical models for varying constants
mentioned above. Motivated by theoretical predictions and
observational data, we will study the formation of the light elements
in the early universe assuming that the values of the gauge coupling
constants of the fundamental interactions (electromagnetic, strong and
weak) may be different from their actual value. Thus, our approach is
a phenomenological one and our results will be model
independent. Furthermore, we assume the chiral limit of QCD to analyze
the dependence of nucleon masses, binding energies and cross sections
with the strong interaction coupling constants.  The gauge coupling
constants of $U(1)$, $SU(2)$ and $SU(3)$, namely, $\alpha_1$,
$\alpha_2$ and $\alpha_3$ are related with the fine structure constant
$ \alpha $, the QCD energy scale $\Lambda _{QCD}$ and the Fermi
coupling constant $G_F$ through the following equations:
\begin{equation}
\alpha ^{-1}\left( E\right) =\frac 52\alpha _1^{-1}\left( E\right) +\alpha
_2^{-1}\left( E\right) \label{equ:alpha:12}
\end{equation}
\begin{equation}
\Lambda _{QCD} =E\exp \left[ -\frac{2\pi }7\alpha
_3^{-1}\left( E\right) \right] \label{equ:LQCD:3}
\end{equation}
\begin{equation}
G_F=\frac{\pi \ \alpha _2\left( M_Z\right) }{\sqrt{2}M_Z^2}
\label{equ:GF:2} 
\end{equation}
where $E$ refers to the energy scale and $M_Z$ refers to the boson
$Z$ mass. Actually, we will study the dependence of the different
physical quantities involved in the calculation of the primordial
abundances with $\alpha$, $\Lambda_{QCD}$ and $G_F$. 

Almost all of the observational and experimental data are consistent
with no variation of the constants \citep{LV02}. Moreover, the
reported variations \citep{Murphy03b,Ivanchik03} are very small
($\frac{\Delta \alpha_i}{\alpha_i} \sim 10^{-5}$). Therefore, in order
to find out the dependences of relevant physical quantities with
$\alpha$, $\Lambda_{QCD}$ and $G_F$ we will perform a Taylor expansion
to first order in each case as follows:
\begin{eqnarray}
\Delta Q &=&\frac {\partial Q}{\partial
\alpha}\arrowvert_{(\alpha^{today}, \Lambda_{QCD}^{today}, G_F^{today})} \Delta
\alpha  \nonumber \\ 
&&+\frac {\partial Q}{\partial
\Lambda_{QCD}}\arrowvert_{(\alpha^{today}, \Lambda_{QCD}^{today}, G_F^{today})}
\Delta \Lambda_{QCD}  
+\frac {\partial Q}{\partial
G_F}\arrowvert_{(\alpha^{today}, \Lambda_{QCD}^{today},  G_F^{today})}
\Delta G_F
\end{eqnarray}
where $Q$ refers to the physical quantities involved in the
nucleosynthesis calculation such as nucleon and nucleus masses,
nuclear binding energies, cross sections and abundances of the
elements.

In the standard picture, the only free parameter of the
nucleosynthesis calculation is the density of baryonic matter
$\Omega_B h^2$. This quantity has been determined with a great
accuracy with data from the CMB provided by WMAP \citep{wmapest}. On
the other hand, the baryon density can also be estimated using data
provided by galaxy surveys (SDSS, 2dF) and x-ray satellites (Chandra,
XMM-Newton, ROSAT, ASCA).  In appendix \ref{rayos} we combine
different data to obtain an estimation of $\Omega_B h^2$ independent
of the WMAP estimate. Therefore, we shall approach to the problem
studying the dependences of all physical quantities and abundances
with both the fundamental constants and the baryon density. Thus, we
will obtain the uncertainties of the abundances of the light elements
as function of the variations of the fundamental constants with
respect to their actual value and as function of the variation of
$\Omega_B h^2$ with respect to the WMAP estimate
\citep{wmapest}. On the other hand, we will also compare the predicted theoretical
expressions for the abundances with observational data and include
independent estimates of the baryon density in the analysis (see
section \ref{analisis}).

Furthermore, in section \ref{masas}, we shall calculate the dependence
of the nucleon masses and binding energies with the fundamental
constants, and in section \ref{scattering}, the corresponding
dependence of the relevant scattering cross sections. We have carried
this calculations in some detail, since there are several subtle
points in these dependences that will be clearly exhibited in the final
results. In section
\ref{abundancias} we apply the semi-analytical method proposed by
\citet{Esma91} to calculate the abundances of the light elements and
their dependence with the fundamental constants. In section \ref{analisis}
we briefly describe the observational data and the results of
comparing them with the theoretical predictions calculated in this
work. We also discuss our conclusions.


\section{Masses and binding energies of light elements}
\label{masas}

In this section we analyze the dependence of the nucleon masses,
nuclear binding energies and nuclei masses of the light elements with
the fundamental constants $\alpha$ and $\Lambda_{QCD}$. The weak
interaction contribution is too small to produce any observable
consequences \citep{HW76,CV02}.

The dependence of the hadronic masses and nuclear binding energies
with the QCD coupling constant $\alpha_3$ or the QCD scale parameter
$\Lambda_{QCD}$ depends on the model of hadronic interactions
considered. However, if we assume that the quark masses are null, an
assumption which is called in the literature as chiral limit, there is
a only a single parameter in the theory, namely the QCD scale
parameter $\Lambda_{QCD}$. Even though great efforts
\citep{BS03,BS03b,EMG03,Flambaum02,Flambaum03,Flambaum03b,Flambaum04,Olive02} have been
done in order to analyze the dependence of nucleon masses and binding
energies with $\Lambda_{QCD}$ beyond the chiral limit, this task is
not trivial and highly model dependent.

On the other hand, from simple dimensional analysis  \citep{Stev81},
 it follows that in a theory with only one relevant parameter all
static observables with dimension of mass must be proportional to this
parameter, which in our case is $\Lambda_{QCD}$. More precisely, any
quantity $\sigma$ with units of $E^D$ (where $E$ means energy) must
satisfy an equation of the form:
\begin{equation}
\sigma = \Lambda_{QCD}^D f\left[\frac{Q}{ \Lambda_{QCD}}\right]
\end{equation}
where $Q$ is a quantity specifying the energy scale. Furthermore, for
static quantities such as nucleon masses the previous equation takes
the form:
\begin{equation}
\sigma = \Lambda_{QCD}^D f\left[\frac{\sigma}{ \Lambda_{QCD}}\right]
\label{quiral2}  
\end{equation}
since the only scale parameter is $\sigma$ itself. The solution of
equation \ref{quiral2} reads:
\begin{equation}
\sigma = \Lambda_{QCD}^D X \label{lquiral}  
\end{equation}
where $X$ is a dimensionless numerical constant. In such way, all
low-energy static quantities will satisfy an equation of the form
\ref{lquiral}. Moreover, all nucleon masses and energies will have a
linear dependence:
\begin{equation}
m_N \sim \epsilon_B \sim \Lambda_{QCD}  
\end{equation}
and all nuclear radii will satisfy:
\begin{equation}
R \sim \Lambda_{QCD}^{-1}  
\end{equation}
since we use units where $\hbar = c = 1$ for this analysis. The chiral  limit
was previously considered by \citet{SV90} studying time variation of
fundamental constants in planetary phenomena.

The mass of the  nucleons can be written as a sum of two
contributions: the electromagnetic contribution  $m_N^C$ and the strong
interaction contribution $m_N^S$:
\begin{eqnarray}
m_N=m_N^C+m_N^S
\end{eqnarray}
The
electromagnetic contribution depends on the nuclear radius $R$ as follows:
\begin{equation}
\epsilon_C=\frac{Z}{4 \pi \epsilon_0} \frac{e^2}{R}
\label{ecoulomb}
\end{equation}
Therefore the electromagnetic contribution to the nucleon mass in the
chiral limit has the following dependence with $\Lambda_{QCD}$:
\begin{equation}
 m_N^C \sim \Lambda_{QCD}
\end{equation}

\citet{Cot63} used perturbation theory to calculate the
electromagnetic self energy of a nucleon $m_N$ to first order in
$\alpha$:
\begin{equation}
m_N^C\sim K \alpha 
\end{equation}
where $K$ can be expressed as a function of Sachs form factors
$G^N_{E,M}$, which can be calculated from measurements of
electron-nucleon scattering. On the other hand, the strong
interaction contribution to the mass in the chiral limit is proportional to
$\Lambda_{QCD}$. Therefore, we can write:
\begin{eqnarray}
m_N^C &=& m_N^C \frac{\alpha}{\alpha^{today}}
\frac{\Lambda_{QCD}}{\Lambda_{QCD}^{today}}\label{m1}\\ 
m_N^S &=& m_N^S\frac{\Lambda_{QCD}}{\Lambda_{QCD}^{today}}\label{m2}
\end{eqnarray}
%
After performing a Taylor expansion to first order, as explained in
section \ref{sec:Intro} and using equations \ref{m1} and \ref{m2}, we
obtain the dependence of the nucleon masses with the fundamental
constants:
\begin{equation}
  \frac{\delta m_N}{m_N}=\frac{m_N^C}{m_N} \frac{\delta
\alpha}{\alpha}+\frac{\delta \Lambda_{QCD}}{\Lambda_{QCD}} = P \frac{\delta
\alpha}{\alpha}+\frac{\delta \Lambda_{QCD}}{\Lambda_{QCD}}
\end{equation}
The values of $P$ are shown in table \ref{resultados1}. 

\begin{table}[h]
\caption{Dependence of nucleon and nuclei masses with the fundamental
constants:$\frac{\delta m_N}{m_N}= P \frac{\delta
\alpha}{\alpha}+\frac{\delta \Lambda_{QCD}}{\Lambda_{QCD}}$}
\label{resultados1}
\begin{center}
\begin{tabular}{|c|c|}
\hline

Nucleon/Nucleus & $P (\times 10^{-4})$\\ \hline

$m_p$&6.71\\\hline

$m_n$&-1.38\\\hline

$\De$&2.67\\\hline

$\Tr$&1.32\\\hline

$\Het$&1.05\\\hline

\He&0.66\\ \hline

\Lis&1.50\\\hline

\Li&1.14\\\hline

$^7{\rm Be}$&2.30\\\hline
\end{tabular}
\end{center}
\end{table}





Next, we analyze the dependence of the nuclei masses with $\alpha$ and
$\Lambda_{QCD}$. As we did for nucleons, we perform a Taylor expansion
to first order to obtain for a nucleus of mass $m_x$ the following
expression:
\begin{equation}
\label{derivada-de-masa}
 \frac{\delta
m_x}{m_x}=(A-Z)\frac{m_n}{m_x}\frac{\delta
m_n}{m_n}+Z\frac{m_p}{m_x}\frac{\delta
m_p}{m_p}-\frac{\epsilon_x}{m_x}\frac{\delta
\epsilon_x}{\epsilon_x}  
\end{equation}
In the more general case, the binding energy $(\epsilon_x)$ can be
written as a sum of two terms: the electromagnetic contribution
($\epsilon_C$) and the strong interaction contribution ($\epsilon_S$)
as follows: $\epsilon_x = \epsilon_C+\epsilon_S$. However, in the
cases of nuclei with only one proton ($\De$ and $\Tr$), there is no
electromagnetic interaction and therefore the electromagnetic
contribution ($\epsilon_C$) is null . On the other hand, the same
arguments that were used to obtain equations \ref{m1} and \ref{m2} can
be applied for the binding energy to obtain:
\begin{equation}
\label{energia} \frac{\delta
\epsilon_x}{\epsilon_x}=\frac{\epsilon_C}{\epsilon_x} \frac{\delta
\alpha}{\alpha}+ \frac{\delta \Lambda_{QCD}}{\Lambda_{QCD}}  
\end{equation}
Inserting this last expression in equation \ref{derivada-de-masa}, we
obtain the general expression for the dependence of a nucleus mass
with $\alpha$ and $\Lambda_{QCD}$:
\begin{eqnarray}
\label{deltamasas2} \frac{\delta m_x}{m_x}&=&
P \frac{\delta \alpha}{\alpha}+ \frac{\delta
\Lambda_{QCD}}{\Lambda_{QCD}}
\end{eqnarray}
The values  $P$ for different nuclei  are shown in table \ref{resultados1}.

\section{Thermonuclear reaction rates}
\label{scattering}

In this section we calculate the thermonuclear reaction rates as
functions of fundamental constants.  We
also show the dependence of the reaction rates with the baryon density
$\rho_B=\Omega_B h^2$. Following \citet{Esma91} we can write the
thermonuclear reaction rate as:
\begin{equation}
\label{tasa} 
[ij\rightarrow kl]=\rho_B N_A \langle\sigma v\rangle=0.93 \times
10^{-3} \Omega_B h^2 T^3_9 N_A \langle\sigma v\rangle \frac{1}{\textrm{seg}}  
\end{equation}
where $\sigma$ is the cross section, $v$ is the relative velocity,
$\rho_B=0.93 \times 10^{-3} \Omega_B h^2 T^3_9
\frac{\textrm{g}}{\textrm{cm}^3}$ is the density of baryonic matter,
$N_A$ is Avogadro's number per gram, $T_9$ is the temperature in units
of $10^9 K$. 

Using a Maxwell-Boltzmann distribution in velocities, the Boltzmann
averaged cross section, $\langle\sigma v\rangle$ can be expressed as
follows:
\begin{equation}
\label{sv2} \langle\sigma v\rangle=\left( \frac{\mu}{2 \pi kT} \right) ^{3/2}
\int e^{-\frac{\mu v^2}{2kT}}v \sigma(E)d^3v  
\end{equation}

We need to find out $\frac{\delta [ij\rightarrow kl]}{[ij\rightarrow
kl]}$ as a function of the relative variations of the fundamental
constants, $\left(\frac{\delta \alpha}{\alpha}, \frac{\delta
\Lambda_{QCD}}{\Lambda_{QCD}}, \frac{\delta G_F}{G_F} \right) $ and
$\frac{\delta \Omega_B h^2}{\Omega_B h^2}$: the relative variation of
the value of the baryon density with respect to WMAP estimate
$\Omega_B h^2 =0.0224$  \citep{wmapest}.  The
temperature does not depend on the values of the fundamental
constants, but the final temperature of each stage does and therefore,
we can write:
\begin{equation}
 \label{derivada} \frac{\delta [ij\rightarrow kl]}{[ij\rightarrow
kl]}=
\frac{\delta \Omega_B h^2}{\Omega_B h^2}+
3 \frac{\delta T_9^f}{T_9^f}+\frac{\delta \langle\sigma v\rangle}{\langle\sigma v\rangle} 
\end{equation}
where $T_9^f=f(\alpha,\Lambda_{QCD}, G_F)$ for all the reaction
rates. On the other hand, $\frac{\delta \langle\sigma
v\rangle}{\langle\sigma v\rangle}$ depends on the fundamentals
constants through the masses of the nucleons and light nuclei and the
form factor of the reactions. In the general case, there are not
analytic expressions for $\sigma(E)$ derived from ``first
principles''. We suggest several expressions that attempt to fit
$\sigma(E)$, according to the elements in the reactions .


\subsection{Cross sections for charged particles reactions}

The cross section for charged particles reactions  is given by
\cite{Fowler1,Fowler2,Wagoner}:
\begin{eqnarray}
\label{senr} \sigma = \frac{S(E)}{E} e^{-2 \pi \alpha Z_1 Z_2
\sqrt{\mu c^2 /2E}}
\end{eqnarray}
where $Z_i$ is the charge of the $i$ particle, $\mu=\frac{m_1
m_2}{m_1+m_2}$ is the reduced mass, $E$ is the energy, $S(E)$ is the
form factor. The dependence of the cross sections for charged particle
reactions have been analyzed previously \citep{Iguri99,Nollet}. In
particular, \citet{Nollet} improved the analysis and studied the form
factor as a function of $\alpha$. In this paper, we use the criteria
established by these authors to analyze the dependence of the form
factor with $\alpha$.

Next, we analyze the dependence of the form factor with
$\Lambda_{QCD}$ using dimensional arguments and the chiral limit. The
units of the cross section are ${\rm cm}^2$ and therefore it follows
that in a theory with massless quarks $\sigma \sim
\Lambda_{QCD}^{-2}$. The only quantity of eq. \ref{senr} that has
units is the factor $\frac{S(E)}{E}$ and thus we obtain:
\begin{equation}
S(E)\sim \Lambda_{QCD}^{-1}  
\end{equation}

This is valid for all charged particle reactions. The exact dependence
of the form factor $S(E)$ with the energy is unknown. However, as it
is usually done in the literature \citep{Fowler1,Fowler2,Wagoner}, we
can do a MacLaurin expansion:
\begin{equation}
\label{s} S(E)=S(0)\left(1+\left(\frac{dS}{dE}\right)_{E=0}
\frac{1}{S(0)}E+ \frac{1}{2} \left(\frac{d^2S}{dE^2}\right)_{E=0}
\frac{1}{S(0)}E^2\right)  
\end{equation}
where $\frac{dS}{dE}$ and $\frac{d^2S}{dE^2}$ are expressed in {\rm
barn} and {\rm barn Mev}$^{-1}$ respectively. The terms inside the
brackets have no dimensions, therefore:
\begin{equation}
S(0)\sim \Lambda_{QCD}^{-1}  
\end{equation}
The dependence of the charged particle cross sections with $\alpha$
has been analyzed by \citet{Nollet}, yielding:
\begin{equation}
S(0) \sim \alpha  
\end{equation}
Furthermore, it follows that all radiative capture rates should be
multiplied by a factor $\frac{\alpha}{\alpha^{today}}$, except the
reactions $\Tr(\alpha \gamma) \Li$ and $\Het (\alpha \gamma) \Be$. This
cross sections should be multiplied by $f(\alpha)=\sum b_i
\left[\frac{\alpha} {\alpha^{today}}-1 \right]$ (see table
\ref{correccion2}). Finally, in the cases in which the reaction
produces two charged particles, the cross section should by multiplied
by $1-b+b\frac{\alpha}{\alpha^{today}}$ (see table \ref{correcciones}
).

We insert the expression for $\sigma(E)$ into the equation
(\ref{sv2}) in order to calculate the  Boltzmann averaged cross
sections:
\begin{equation}
\label{sigv} \langle\sigma v\rangle = \sqrt{\frac{8}{\mu \pi}}
(kT)^{-1/2}\sum_{i=0}^{2} \frac{(kT)^i}{i!}
\left(\frac{d^iS}{dE^i}\right)_{E=0} \int_0^{\infty} y^i e^{-y-\xi
y^{-1/2}} dy  
\end{equation}
where $\xi= 2 \pi \alpha Z_1 Z_2 \sqrt{\mu c^2 /2kT}$ and the masses
in kg. The integrals are calculated in \citet{Iguri99}. Tables
\ref{cargadas1} and \ref{cargadas2} show the dependence of charged
particles reaction rates with the fundamental constants.

\begin{table}
\caption{Radiative captures, its dependence on $\alpha$ $\left (
f(\alpha)=\sum b_i \left[\frac{\alpha}{\alpha^{today}}-1
\right]\right) $} \label{correccion2}
\begin{center}
\begin{tabular}{|c|c|c|c|c|c|c|}
\hline Reaction &$b_0$&$b_1$&$b_2$&$b_3$&$b_4$&$b_5$ \\ \hline
$\T(\alpha,\gamma)\Li$&1&1.372&0.502&0.183&0.269&-0.218\\
\hline
$\Het(\alpha,\gamma)\Be$&1&2.148&0.669&-5.566&-10.630&-5.730 \\
\hline
\end{tabular}
\end{center}
\end{table}

\begin{table}
\caption{Dependence on $\alpha$ of different kinds of reactions
rates} \label{correcciones}
\begin{center}
\begin{tabular}{|c|c|}
\hline Reaction & Multiplied by \\ \hline

Charged particles reaction rates &$\frac{\alpha}{\alpha^{today}}$
\\ \hline

Photon emission & $\frac{\alpha}{\alpha^{today}}$
\\ \hline

$\D(d,p)$ & $1+0.16-0.16\frac{\alpha}{\alpha^{today}}$  \\ \hline
$\Het(n,p)$& $1-0.30+0.30\frac{\alpha}{\alpha^{today}}$ \\ \hline
$\Het(d,p)$& $1+0.09-0.09\frac{\alpha}{\alpha^{today}}$ \\ \hline
$\Li(p,\alpha)$& $1+0.18-0.18\frac{\alpha}{\alpha^{today}}$
\\ \hline $\Be(n,p)$& $1-0.20+0.20\frac{\alpha}{\alpha^{today}}$
\\\hline
\end{tabular}
\end{center}
\end{table}

%

\subsubsection{Cutoff factor}

The truncated MacLaurin series we have use for $S(E)$ diverges at high
energy. Thus, it is important to include a cutoff factor for
non-resonant reaction rates so that they can be used at any energy.
The next term in the expansion for $S(E)$ would be proportional to
$E^3 \sim T^2$, so as it is proposed in the literature
\citep{Fowler1,Fowler2} we consider a cutoff factor: $f_{co}=e^{-(T_9
/T_{co})^2}$, where $T_{co} \sim \frac{E_r}{\alpha}$ and $E_r$ is the
resonant energy \citep{Fowler1,Fowler2}. Therefore we multiply the
expression (\ref{sigv}) by a factor:
\begin{eqnarray}
\label{corte} f_{co}=e^{-\left(\frac{\alpha T_9}{E_r}\right)^2}
\end{eqnarray}
This correction is relevant for the following reactions: $\Het (d,p)\He $, $\T (d,n) \He $, $\Lis (p,\alpha) \T $, $\Lis (\alpha,p) ^{10}{\rm Be}$, $\Li(p,\alpha ) \He $.



\subsubsection{Alternative expression for the form factor} \label{s diferente}

In the MeV range the cross section form factor varies considerably. In
this range the truncated MacLaurin series is not satisfactory so that
it is convenient \citep{Fowler1,Fowler2} to use for $S(E)$ an
expression of the form: $\label{exp} S(E)=S(0) e^{-aE}$. In such way,
the cross sections are given by:
\begin{equation}
\label{sigma raro} \sigma(E)= \frac{S(0)}{E} e^{-aE} e^{-2 \pi
\alpha Z_1 Z_2 \sqrt{\mu c^2 /2E}}  
\end{equation}
where $a$ has no dependence on the fundamental constants. The
quantities with units in equation \ref{sigma raro} are $S(0)$ and $E$,
therefore, in the chiral limit we have $ S(0)
\sim\Lambda_{QCD}^{-1}$. In such way, the Boltzmann cross section
(eq.(\ref{sv2})) yields:
\begin{equation}
\langle\sigma v\rangle = \frac{8}{\sqrt{6}} \left(\frac{\mu}{kT}\right)^{3/2}
\frac{ S(0) kT}{akT+1}\left(\frac{\xi_a^2}{4}\right)^{1/6}
e^{-3\left(\frac{\xi_a^2}{4}\right)^{1/3}}
\left[1+\frac{5}{36}\left(\frac{\xi_a^2}{4}\right)^{-1/3}\right]\frac{\textrm{cm}^3}{\textrm{seg}}  
\end{equation}
where $\xi_a=\xi \sqrt{akT+1}=2 \pi \alpha Z_1 Z_2 \sqrt{\frac{\mu
c^2}{2kT}(akT+1)}$. This alternative expression for non-resonant
reaction rates is relevant for the following reactions: $\Lis
(p,\gamma)\Be$, $\T(\alpha,\gamma)\Li$ and $\Het (\alpha,\gamma)\Be$


\subsection{Resonant charged particle reaction rates}
\label{resonancia}

The expressions for the cross sections vary with the temperature.
Moreover, in the range of energies relevant for our calculation there
are certain reactions that proceed through many resonances. In this
case, we have to include an extra term in the cross section. There are
two kinds of resonances: i) Single Resonance, ii) Continuum Resonance. 
In each case we use the expressions given by \citet{Fowler1,Fowler2}.

\subsubsection{Resonance cross sections}

In this case, the following expression provides a good fit to the
cross section: \cite{Fowler2}:
\begin{equation}
\label{resonancia1} \sigma(E)=\frac{\pi \hbar^2}{2 \mu E}
\frac{\omega_r \Gamma_1 \Gamma_2}{(E-E_r)^2+\Gamma^2/4}  
\end{equation}
where $\Gamma_i$ is the partial with for the decay of the resonant
state by the reemission of $(i-1)+i$, $\Gamma$ is the sum over all
partial widths (the partial widths are not functions of  $\alpha$),
$\omega_r=\frac{\left(1+\delta_{ab}\right) g_r}{g_a g_b}$ and $g_r=2
J_r+1$, $J_r$ being the spin of the resonant state, $\mu$ in kg and
$E_r$ is the resonance energy in the center of momentum system and
depends on the nuclear radius.  Finally, the Boltzmann cross section
$\langle\sigma v\rangle$ can be calculated as follows:
\begin{equation}
\langle\sigma v\rangle = \left(\frac{2 \pi \hbar^2}{\mu kT}\right)^{3/2}
\frac{(\omega \gamma)_r}{\hbar} e^{E_r/kT} \hskip 0.5cm
\frac{\textrm{cm}^3}{\textrm{seg}}  
\end{equation}
where $\gamma_r=\left(\frac{\Gamma_1 \Gamma_2}{\Gamma}\right)_r$.
Here, the cross section depends on the fundamental constants through the
final temperature and the resonance energy.  Besides, the resonance width is also a function of the fundamental constants, but the cross section is much less sensitive to this dependence. 

In a theory with massless
quarks: $E_r \sim \Lambda_{QCD}$. On the other hand, $E_r$ does not
depend on $\alpha$. The dependence of the temperature will be analyzed
in section \ref{abundancias}.  This correction is relevant for the
following reactions: $\D (\alpha,\gamma) \Lis$, $\Lis (p,\alpha)\T$,
$\Lis(\alpha,\gamma)^{10}{\rm Be}$, $\Be(p,\gamma) ^8{\rm Be}$ and $\Li (p,\alpha)\He$.

\subsubsection{Continuum resonances}

When the temperature scale is of order $T_9 \sim 1$ , there are
several reactions that proceed through many resonances that are
separated by intervals not greater than their widths or that overlap
to form a continuum. In this cases, the cross section can be written
as \cite{Fowler2}:
\begin{eqnarray}
\sigma(E)=\left \{
\begin{array}{cc}
 2\sigma(2C)\frac{C}{E}\left(\frac{E}{C}-1\right)^{m+1/2}
& \textrm{si}\hskip 0.5cm E\geq C   \\
    0& \textrm{si} \hskip 0.5cm E\leq C
  \end{array}\right.
\end{eqnarray}
where $m$ is integer or rational fraction, $C$ is the effective
continuum threshold energy and $\sigma(2C)$ is the cross section at
$E=2C$. 

After inserting this expression in the integral (\ref{sv2}),
we obtain:
%
\begin{equation}
\langle\sigma v\rangle = \Gamma(m+3/2) \sigma(2C) \sqrt{\frac{32 C}{\pi
\mu}}\left(\frac{kT}{C}\right)^m e^{-C/kT}  
\end{equation}
where $\Gamma(m+3/2)$ is the gamma function. On the other
hand, $C$ has units of energy. Therefore, in the chiral limit $C \sim
\Lambda_{QCD}$.  This correction is relevant for the following
reactions: $\Het(d, p)\He$, $\T (d,n)\He$, $\Lis(p,\alpha)\T$
and $\Lis(\alpha,\gamma) ^{10}{\rm Be}$.


\begin{table}[h]
\caption{Charged particles reaction rates $\Theta=\Omega_B h^2
T_9^{7/3} {\alpha}^{1/3} {\mu}^{-1/3}$,  $\Psi={\mu \alpha^2}$,
$\Xi(b)=\Omega_B h^2 \mu^{-b}$, \hskip 0.5cm
$\Sigma(a)=\frac{\alpha}{\alpha^{today}}\left(1+a-a\frac{\alpha}{\alpha^{today}}\right)$,
\hskip 0.5cm $P_{IB}(\Psi, T_9, c_1, c_2, c_3, c_4, c_5)=
1+c_1\times 10^{-12} \hskip 0.05cm \Psi^{-1/3} \hskip 0.05cm
T_9^{1/3}+c_2 \times 10^{10} \hskip 0.05cm \Psi^{1/3}\hskip 0.05cm
T_9^{2/3}+c_3 \hskip 0.05cm T_9+ c_4 \times 10^{20}\hskip 0.05cm
\Psi^{2/3}\hskip 0.05cm T_9^{4/3}+c_5 \times 10^{10}\hskip 0.05cm
\Psi^{1/3} \hskip 0.05cm T_9^{5/3}$} \label{cargadas1}
\begin{center}
\begin{tabular}{|c|l|}
\hline

Reaction & Reaction rate $\left(\frac{1}{\rm seg}\right)$\\ \hline

$^3\rm {H} \left( \rm {p}, \gamma \right) ^4\rm {He}$ & $1.14
\times 10^{-7}\hskip 0.2cm \Theta \hskip 0.2cm
\left[\Sigma(0)\right]^2 e^{-9.55 \times
10^{10}\left(\frac{\Psi}{T_9}\right)^{1/3}} \times$\\
 & \hskip 0.3cm $P_{IB}\left(\Psi,\hskip 0.1cm  T_9,\hskip 0.1cm 4.36,\hskip 0.1cm 4.14,\hskip 0.1cm 1.26,\hskip 0.1cm 3.35,\hskip 0.1cm 2.61\right) $\\
\hline

$^2\rm {H} \left (d, \rm {n}\right) ^3\rm {He}$& $2.26 \times
10^{-3}\hskip 0.2cm \Theta \hskip 0.2cm \Sigma(0)\hskip 0.2cm
e^{-9.55 \times 10^{10}\left(\frac{\Psi}{T_9}\right)^{1/3}}
\times$ \\
 &  \hskip 0.3cm $P_{IB}\left(\Psi,\hskip 0.1cm  T_9,\hskip 0.1cm 4.36, \hskip 0.1cm 1.96, \hskip 0.1cm 0.6, \hskip 0.1cm -0.206, \hskip 0.1cm -0.16\right)$\\\hline

$^2\rm {H} \left( ^3\rm {He}, \rm{p} \right) ^4\rm {He}$&
$0.39\hskip 0.2cm \Theta \hskip 0.2cm \Sigma(0.09) \hskip 0.2cm
e^{-1.52\times 10^{11}\left(\frac{\Psi}{T_9}\right)^{1/3}-(507.36
T_9 \alpha)^2} \times$\\ & \hskip 0.3cm $P_{IB}\left(\Psi, \hskip
0.1cm T_9,\hskip 0.1cm 2.75,\hskip 0.1cm -2.16,\hskip 0.1cm
-0.42,\hskip 0.1cm 13.5, \hskip 0.1cm 6.58\right)+$\\
 &$+2.63 \times 10^{-8}\hskip 0.2cm \Xi\left( \frac{1}{2}\right) \hskip 0.2cm \Sigma(0.09)\hskip 0.2cm T_9^{5/2} \hskip 0.2cm e^{-1.76 T_9^{-1}} $\\\hline

$^3\rm {H}\left( \rm{d}, \rm {n}\right) ^4\rm {He}$& $0.49 \hskip
0.2cm \Theta\hskip 0.2cm \Sigma(0)\hskip 0.2cm e^{-9.55\times
10^{10}\left(\frac{\Psi}{T_9}\right)^{1/3}-(1141.67 T_9 \alpha)^2}
\times$\\ & \hskip 0.3cm  $P_{IB}\left(\Psi,\hskip 0.1cm
T_9,\hskip 0.1cm 4.36,\hskip 0.1cm 3.78,\hskip 0.1cm 1.16,\hskip
0.1cm 46.8,\hskip 0.1cm 3.64 \times 10^{11}\right)+$\\
 & $+3.39 \times 10^{-8}\hskip 0.2cm  \Xi\left(\frac{1}{2}\right) \hskip 0.2cm \Sigma(0)\hskip 0.2cm T_9^{7/3} \hskip 0.2cme^{-0.523 T_9^{-1}} $\\\hline

$^2\rm {H}\left(\rm{d}, \rm{p}\right)^3\rm {H}$&$2.37 \times
10^{-3}\hskip 0.2cm \Theta\hskip 0.2cm \Sigma(0.16)\hskip 0.2cm
e^{-9.55\times 10^{10}\left(\frac{\Psi}{T_9}\right)^{1/3}}
\times$\\ & \hskip 0.3cm $P_{IB}\left(\Psi,\hskip 0.1cm T_9,\hskip
0.1cm 4.36,\hskip 0.1cm 1.16,\hskip 0.1cm 0.35,\hskip 0.1cm
-0.051,\hskip 0.1cm -0.04\right)$\\\hline

$^2\rm {H} \left(\alpha, \gamma \right) ^6\rm {Li}$& $1.88 \times
10^{-10}\hskip 0.2cm \Theta\hskip 0.2cm  \left[\Sigma(0)\right]^2
\hskip 0.2cm e^{-1.52\times
10^{11}\left(\frac{\Psi}{T_9}\right)^{1/3}} \times$\\
 &  \hskip 0.3cm $P_{IB}\left(\Psi,\hskip 0.1cm  T_9,\hskip 0.1cm 2.75,\hskip 0.1cm -9.9,\hskip 0.1cm 8.85,\hskip 0.1cm  -2.43,\hskip 0.1cm -1.19\right)+$\\
 & $+8.27 \times 10^{-39} \hskip 0.2cm \Xi\left(\frac{3}{2}\right) \hskip 0.2cm\left[\Sigma(0)\right]^2 \hskip 0.2cm T_9^{3/2} \hskip 0.2cm e^{-8.228  T_9^{-1}}$\\\hline

$\rm {H} \left(^6\rm {Li}, \alpha \right) ^3\rm {H}$&$0.20\hskip
0.2cm \Theta \hskip 0.2cm \Sigma(0)\hskip 0.2cm e^{-1.99\times
10^{11}\left(\frac{\Psi}{T_9}\right)^{1/3}-(24.94 T_9 \alpha)^2}
\times$\\ &  \hskip 0.3cm $P_{IB}\left(\Psi, \hskip 0.1cm
T_9,\hskip 0.1cm 2.10,\hskip 0.1cm -0.14, \hskip 0.1cm - 0.02,
\hskip 0.1cm  0.033, \hskip 0.1cm 0.012 \right)+$\\ & $+4.53
\times 10^{-8} \hskip 0.2cm \Xi\left(\frac{1}{2}\right) \hskip
0.2cm \Sigma(0) \hskip 0.2cm T_9^2 \hskip 0.2cm e^{-21.82
T_9^{-1}}+$\\ & $+ 6.68 \times 10^{-34}\hskip 0.2cm
\Xi\left(\frac{3}{2}\right) \hskip 0.2cm  \Sigma(0)\hskip 0.2cm
T_9^{3/2}\hskip 0.2cm e^{-17.76  T_9^{-1}}$\\\hline
\end{tabular}
\end{center}
\end{table}

\begin{table}[h]
\caption{Charged particles reaction rates $\Theta=\Omega_B h^2
T_9^{7/3} {\alpha}^{1/3} {\mu}^{-1/3}$,  $\Psi={\mu \alpha^2}$,
$\Xi(b)=\Omega_B h^2 \mu^{-b}$, \hskip 0.5cm
$\Sigma(a)=\frac{\alpha}{\alpha^{today}}
\left(1+a-a\frac{\alpha}{\alpha^{today}} \right)$,\hskip 0.5cm
$P_{IB}(\Psi, T_9, c_1, c_2, c_3, c_4, c_5)= 1+c_1\times 10^{-12}
\hskip 0.05cm \Psi^{-1/3} \hskip 0.05cm T_9^{1/3}+c_2 \times
10^{10} \hskip 0.05cm \Psi^{1/3}\hskip 0.05cm T_9^{2/3}+c_3 \hskip
0.05cm T_9+ c_4 \times 10^{20}\hskip 0.05cm \Psi^{2/3}\hskip
0.05cm T_9^{4/3}+c_5 \times 10^{10}\hskip 0.05cm \Psi^{1/3} \hskip
0.05cm T_9^{5/3}$, \hskip 5cm $P_{LN}(x,d_1,d_2,d_3,d_4,d_5)=
1+d_1 \hskip 0.05cm x+d_2 \hskip 0.05cm x^2+d_3 \hskip 0.05cm
x^3+d_4 \hskip 0.05cm x^4 + d_5 \hskip 0.05cm x^5$, $T_{9v}={v T_9
+11.605}$ } \label{cargadas2}
\begin{center}
\begin{tabular}{|c|l|}
\hline Reaction & Reaction rate $\left(\frac{1}{\rm seg}\right)$\\
\hline

$^3\rm {H}\left(\alpha,\gamma\right)^7\rm {Li}$&$\Theta \hskip
0.2cm \left[\Sigma(0)\right]^2 \hskip 0.2cm
P_{LN}\left(\frac{\alpha-\alpha^{today}}{\alpha^{today}},\hskip
0.05cm 3.17,\hskip 0.05cm 0.50,\hskip 0.05cm 0.18,\hskip 0.05cm
0.27,\hskip 0.05cm -0.22\right) \times$ \\ & $ \hskip 0.2cm
\left\{7.47 \times 10^{-6} \hskip 0.2cm e^{-1.52\times
10^{11}\left(\frac{\Psi}{T_9}\right)^{1/3}} \times \right. $\\ &
$\left. \hskip 0.8cm P_{IB}\left(\Psi, \hskip 0.05cm T_9, \hskip
0.05cm 2.75, \hskip 0.05cm -0.76,\hskip 0.05cm -0.15 ,\hskip
0.05cm 0.36,\hskip 0.05cm 0.18\right)+ \right.$ \\ & $\left.
\hskip 0.2 cm +2.68 \times 10^{-5} \hskip 0.2cm T_{9v}^{-5/6}
e^{-6.69\times 10^{10}\left(\frac{\Psi T_{9 v}}{T_9}\right)^{1/3}}
\right\}$ \\ & $v=1.59$ \\ \hline

$^3\rm {He}\left(\alpha, \gamma \right)^7\rm {Be}$&$\Theta \hskip
0.05cm \left[\Sigma(0)\right]^2 \hskip 0.05cm
P_{LN}\left(\frac{\alpha-\alpha^{today}}{\alpha^{today}},\hskip
0.05cm 2.15,\hskip 0.05cm 0.67,\hskip 0.05cm -5.57,\hskip 0.05cm
10.63,\hskip 0.05cm -5.73\right)\times$ \\ & $ \hskip 0.1cm
\left\{3.27 \times 10^{-5} \hskip 0.2cm e^{-2.41\times
10^{11}\left(\frac{\Psi}{T_9}\right)^{1/3}} \times \right.$\\ &
$\left. \hskip 0.3 cm P_{IB}\left(\Psi,T_9, 1.73, -0.0019,
-0.00024,-0.00028, -8.8 \times 10^{-5}\right)+\right.$\\ & $\left.
\hskip 0.1cm +3.12 \times 10^{-4} T_{9v}^{-5/6} e^{-1.06\times
10^{11}\left(\frac{\Psi T_{9v}}{T_9}\right)^{1/3}}\right\}$\\
&$v=1.24$ \\ \hline

$\rm {H}\left(^7\rm {Li}, \alpha\right) ^4\rm {He}$&$3.33 \times
10^{-3} \hskip 0.2cm \Theta \hskip 0.2cm  \Sigma(0) \hskip 0.2cm
e^{-1.99\times
10^{11}\left(\frac{\Psi}{T_9}\right)^{1/3}-\left(0.22 \alpha T_9
\right)^2} \times$\\
 & $\hskip 0.3cm  P_{IB}\left(\Psi, \hskip 0.05cm T_9, \hskip 0.05cm 2.10, \hskip 0.05cm 3.65, \hskip 0.05cm 0.54, \hskip 0.05cm - 5.30, \hskip 0.05cm -1.98\right)+$\\
 & $+ \Xi\left(\frac{2}{3}\right) \hskip 0.2cm \Sigma(0)\hskip 0.2cm
 T_9^{3/2}\times$\\
&$\hskip 0.2cm \left[ 5.54 \times 10^{-34}  e^{-30.44/T_9}+7.98
\times 10^{-38} e^{-4.479/T_9} \right]$\\\hline

$\rm {H}\left(\rm{d}, \gamma \right)^3\rm {He}$&$1.11 \times
10^{-8}\hskip 0.2cm \Theta \hskip 0.2cm  \left[\Sigma(0)\right]^2
\hskip 0.2cm e^{-9.545 \times
10^{10}\left(\frac{\Psi}{T_9}\right)^{1/3}} \times$\\
 & $\hskip 0.2cm P_{IB}\left(\Psi, \hskip 0.05cm T_9, \hskip 0.05cm 4.36, \hskip 0.05cm 8.66, \hskip 0.05cm 2.65, \hskip 0.05cm 1.26, \hskip 0.05cm 0.98\right)$\\\hline

\end{tabular}
\end{center}
\end{table}








\subsection{Non charged particles reaction rates}
\label{no cargadas}

In this case, there is no Coulomb barrier so the cross section cannot
be written as the equation (\ref{senr}). Following \citet{Fowler1} we
write:
\begin{equation}
\sigma(E) = \frac{S(E)}{v}
\end{equation}
where $v$ is the relative velocity.   We consider the expression given by
\citet{Fowler1}:
\begin{equation}
\label{sneutron}
S(E)=S(0)+\left(\frac{dS}{d\widetilde{E}}\right)_{\widetilde{E}=0}
E^{1/2}+ 1/2
\left(\frac{d^2S}{d\widetilde{E}^2}\right)_{\widetilde{E}=0} E  
\end{equation}
where $\widetilde{E}=E^{1/2}$. 

In chiral limit $\sigma \sim \Lambda_{QCD}^{-2}$, and therefore:
\begin{equation}
S(0)\sim \Lambda_{QCD}^{-2}  
\end{equation}

In this way, the expression for the reaction rates (equation
\ref{sv2}) yields:
\begin{equation}
\langle\sigma v\rangle = S(0)\left(1+\frac{2}{\sqrt{\pi}} \frac
{S^{'}(0)}{S(0)}(kT)^{1/2}+\frac{3}{4}\frac {S^{''}(0)}{2S(0)}
kT\right)  
\end{equation}
where $\frac{dS}{d \widetilde{E}}$ is in units of $ \frac{{\rm
cm}^3}{{\rm seg}} {\rm MeV}^{-1/2}$ and $\frac{d^2S}{d \widetilde{E}}$
in in units of $ \frac{{\rm cm}^3}{{\rm seg}} {\rm MeV}^{-1}$

For radiative emission reactions, the cross section should be
multiplied by a factor $\frac{\alpha}{\alpha^{today}}$.  Table \ref{seccion de
neutrones} shows some reaction rates between a neutron and a nucleus.

In some cases the reaction rates of the inverse reactions are
needed. Next, we show the expressions for these reaction rates. For
inverse reactions of the form $[BCAn]$, where neither $B$ or $C$ are
photons, we use the expression given by \citet{Fowler1,Fowler2}:
\begin{equation}
[BCAn]=\frac{2(1+\delta_{BC})g_A}{(1+\delta_{An})g_Bg_C}\left(\frac{m_A 
m_n}{m_B m_C}\right)^{3/2}e^{-Q/kT}
[AnCB]\frac{1}{\textrm{seg}}  
\end{equation}
where $Q=m_A+m_n-m_B-m_C$ . In this case, the form factor
$S(0)$ should be multiplied by $\frac{\alpha}{\alpha^{today}}$ because
of the Coulomb barrier.


For inverse reactions of the form [B$\gamma$nA], we use the expression
given by \citet{Fowler1,Fowler2}:
\begin{equation}
Y_{\gamma}[B\gamma nA]=\frac{g_A
g_n}{(1+\delta_{An})g_B}\left(\frac{m_A
m_n}{m_B}\right)^{3/2}\left(\frac{M_U^2 k T}{2 \pi
\hbar^2}\right)^{3/2} e^{-Q/(kT)} \langle\sigma v\rangle
\frac{1}{\textrm{seg}}  
\end{equation}
where $Q=m_A+m_n-m_B$, $g_n=2 j_n+1=2$ and $M_U=\frac{1}{N_A}$.

In both cases, the additional dependence on the fundamental constants
introduced by the inverse reactions proceed from the temperature (see
section \ref{abundancias}) and the masses (see section \ref{masas}).


\begin{table}[h]
\caption{Non charged particles reactions rates,
$\Sigma_{nc}(a)=1+a-a\frac{\alpha}{\alpha^{today}}$, $P_{F}(T_9,
c_1, c_2)= 1+c_1 \hskip 0.05cm T_9^{1/2}+c_2 \hskip 0.05cm T_9$,
$q=\frac{m_n-m_p}{m_e}$} \label{seccion de neutrones}
\begin{center}
\begin{tabular}{|c|l|}
\hline Reaction & Reaction rate $\left(\frac{1}{\rm seg}\right)$\\
\hline

$\rm {n}\left(\hskip 0.2cm, \rm{e}^-\right)\rm{H}$& $\frac{16}{60}
\pi^2 G_F^2 m_e^5 \left[ \sqrt{q^2-1}\hspace{3pt} (2q^4-9q^2
-8)+\right. $\\ &$\left. \hskip 2.5cm +15 q\hspace{3pt}
\ln\left(q+\sqrt{q^2-1}\right)\right]$
\\ \hline

$\rm {H}\left(\rm {n}, \gamma\right)\rm{d}$ & $40.92\hskip 0.1cm
\Omega_B h^2\hskip 0.1cm \Sigma_{nc}(-1)\hskip 0.1cm T_9^3\hskip
0.1cm P_F(T_9, \hskip 0.05cm -0.86, \hskip 0.05cm 0.43)$
\\ \hline

$Y_{\gamma}\left\{\rm{H}\left(\gamma, \rm {n}\right) \rm {H}
\right\}$ & $2.70 \times 10^{49}\hskip 0.1cm \Sigma_{nc}(-1)\hskip
0.1cm \left(\frac{m_p m_n}{m_d}\right)^{3/2} \hskip 0.1cm
T_9^{3/2}\hskip 0.1cm e^{-11.605 \epsilon_d/T_9}\times $ \\ &
$\hskip 0.2cm  P_F(T_9, \hskip 0.05cm -0.86, \hskip 0.05cm 0.43)$
\\  \hline

$^3\rm {He}\left(\rm {n}, \rm{p}\right) ^3\rm {H}$ &$6.53\times
10^{5}\hskip 0.1cm \Omega_B h^2 \hskip 0.1cm T_9^3 \hskip 0.1cm
\Sigma_{nc}(0.3)\hskip 0.1cm  P_F(T_9, \hskip 0.05 cm -0.59,
\hskip 0.05cm 0.1832)$
\\\hline

$^3\rm {H}\left(\rm{p}, \rm {n}\right)^3\rm {He}$& $6.53\times
10^{5} \hskip 0.1cm \Omega_B h^2\hskip 0.1cm \Sigma_{nc}(0.3)
\hskip 0.1cm \Sigma_{nc}(-1) \hskip 0.1cm \left(\frac{m_3 m_n}{m_T
m_p}\right)^{3/2} \times$\\ & $ \hskip 0.3cm P_F(T_9, \hskip 0.05
cm -0.59, \hskip 0.05cm 0.1832)\hskip 0.1cm e^{-11.605Q_6/T_9}
T_9^{3}$\\ \hline

$^7\rm {Be}\left(\rm {n},\rm{p}\right)^7\rm {Li}$ & $6.27\times
10^{6}\hskip 0.1cm \Omega_B h^2 \hskip 0.1cm  \Sigma_{nc}(0.2)
\hskip 0.1cm T_9^3 \hskip 0.1cm P_F(T_9, \hskip 0.05cm -0.903,
\hskip 0.05cm 0.215 )
$
\\ \hline
\end{tabular}
\end{center}
\end{table}

\subsection{Neutron lifetime}
\label{decaimiento}

Neutron $\beta$ decay is one of the few reactions whose cross section
can be explicitly computed from first principles in terms of the
fundamental constants. It can be approximated by the one point
interaction of neutron, proton, electron and neutrino. The reaction
rate for neutron $\beta$ decay is:
\begin{equation}
n \rightarrow p + e^- +\overline{\nu_e} \label{reaccion}  
\end{equation}

Following \citet{Ichi02} we write the inverse of neutron
lifetime as follows:
\begin{equation}
\frac{1}{\tau}\simeq G_F^2 \int_0^{P_0} d^3 p_e
d^3p_{\nu}\hspace{3pt} \delta(E_e+E_{\nu}+m_p-m_n)  
\end{equation}
where, $G_F$ is Fermi coupling constant, $E_e$ and $E_{\nu}$ are
the electron and neutrino energy, and $p_e$ and $p_{\nu}$ are
the electron and neutrino momenta. After integration we obtain:
\begin{equation}
\label{tau} \frac{1}{\tau}=\frac{16}{60} \pi^2 G_F^2 m_e^5 \left(
\sqrt{q^2-1}\hspace{3pt} (2q^4-9q^2 -8)+15 q\hspace{3pt}
\ln\left(q+\sqrt{q^2-1}\right)\right)
\frac{1}{\textrm{seg}}  
\end{equation}
where $m_e$ is the electron mass,
$q=\frac{Q}{m_e}=\frac{m_n-m_p}{m_e}$


In such way, we obtain the dependence of the neutron decay rate
$\tau^{-1}$ on $G_F$ and on the mass difference (which is a function
of $\alpha$ and $\Lambda_{QCD}$, see section \ref{masas}):
\begin{equation}
\frac{\delta [n]}{[n]}=-\frac{\delta \tau}{\tau}=2 \frac{\delta
G_F}{G_F}+6.54 \frac{\delta \Lambda_{QCD}}{\Lambda_{QCD}}-3.839
\frac{\delta \alpha}{\alpha}  
\end{equation}

\section{Abundances as functions of fundamental constants}
\label{abundancias}

In this section we calculate the abundances of light elements and
their dependence on fundamental constants. First we obtain the neutron
abundance until the freeze-out time of weak interaction. After this
time the neutrons decay freely into protons and electrons, so
their abundance only changes due to this decay.

The general form of the equations that govern the abundances of the
light elements is:
%
\begin{equation}
\label{abgen}
\dot{Y_i}=J(t)- \Gamma(t) Y_i  
\end{equation}
where $J(t)$ and $\Gamma(t)$ are time-dependent source and sink terms
and the dot corresponds to the time derivative. The time-dependent
static solution of this equation is what we will call following
\citet{Esma91} the quasi-static equilibrium (QSE) solution:
%
\begin{equation}
\label{solgn}
f_i=\frac{J(t)}{\Gamma(t)}  
\end{equation}

To determine the formation of light nuclei we shall solve the
following equations using only the most important reactions according
to the rates of production and destruction following the criteria
established by \citet{Esma91}:
\begin{eqnarray}
\label{dotyn} \dot{Y_n}&=&Y_d Y_d [ddn3]+Y_d Y_T [dTn\alpha]+Y_p
Y_T [pTn3]+Y_d Y_{\gamma} [d \gamma np] + \nonumber \\
&&-Y_nY_p[npd\gamma]-Y_nY_3[n3Tp]-Y_n[n]\\ \label{dotyd}
\dot{Y_d}&=&Y_n Y_p [npd\gamma]-2Y_d Y_d \left([ddpT] +[ddn3]
\right)-Y_d Y_T[dTn\alpha]+ \nonumber \\
&&-Y_dY_3[d3p\alpha]-Y_dY_{\gamma}[d\gamma np]-Y_dY_p[dp3\gamma]\\
\label{doty3} \dot{Y_3}&=&Y_d Y_p [pd3\gamma]+Y_T Y_p [pTn3]+ Y_d
Y_d [ddn3]+\nonumber\\ &&-Y_d Y_3 [d3p\alpha]-Y_n Y_3[n3pT]\\
\label{dotyt} \dot{Y_T}&=& Y_n Y_3 [n3pT]+ Y_d Y_d [ddpT] -Y_d Y_T
[dTn\alpha] +\nonumber\\ && -Y_p Y_T [pTn3] -Y_p Y_T [pT \gamma
\alpha] \\ \label{doty6} \dot{Y_6}&=& Y_d Y_{\alpha} [d \alpha 6
\gamma] -Y_n Y_6 [n6\alpha T]-Y_p Y_6 [p6T\alpha]\\ \label{doty7}
\dot{Y_7}&=&Y_nY_{\alpha}[n67\gamma]+Y_nY_B[nBp7]+Y_TY_{\alpha}[T\alpha7\gamma]+\nonumber\\
&&-Y_pY_7[p7\alpha\alpha]-Y_nY_7[n78\gamma]\\ \label{dotyb}
\dot{Y_B}&=&Y_pY_6[p6B\gamma]+Y_3Y_{\alpha}[3\alpha B\gamma]
-Y_{\gamma}Y_B[B\gamma3\alpha]-Y_nY_B[nBp7]+\nonumber\\
&&-Y_pY_B[pB\gamma8]-Y_dY_B[dB\alpha\alpha p]\\ \label{dotyalpha}
\dot{Y_{\alpha}}&=&Y_dY_3[d3p\alpha]+Y_nY_3[n3\alpha\gamma]+Y_dY_T[dTn\alpha]+Y_pY_T[pT\gamma\alpha] 
\end{eqnarray}
where $n$ refers to neutron, $p$ to proton, $d$ to deuterium, $T$ to
tritium, $3$ to $\Het$, $\alpha$ to $\He$, $6$ to $\Lis$,
$7$ to $\Li$, $B$ to $\Be$, $\gamma$ to the photon and $[ijkl]$
is the rate of the reaction $i+j \rightarrow k+l$ and $Y_i$ is the
abundance of the $i$ element relative to baryons $\left(
Y_i=\frac{n_i}{n_B}\right)$. In addition, these equations obey neutron number conservation:
\begin{eqnarray}
\dot {Y_n}+\dot {Y_d}+\dot {Y_3}+2\dot {Y_T}+2\dot
{Y_{\alpha}}=-Y_n [n]
\label{neutroncons}
\end{eqnarray}

The method of \citet{Esma91} consists in calculating the different
abundances between fixed point or stages. We shall solve equations
\ref{dotyn} to \ref{dotyalpha} only for one element in each stage. For
the other elements it is necessary to solve the quasi static
equilibrium equation using only the most important rates of production
and destruction. On the other hand, we perform the calculation of all
final temperatures and abundances and all freeze-out temperatures
numerically. Table \ref{resumen} shows the different stages and the
used equation.

The equations that describe the production of $n$, $\De$, $\Het$ and 
$\Tr$ are independent to the equations for $\Lis$, $\Li$ and
$\Be$. Therefore, we shall solve the first three using the quasi static
equilibrium equation and then we use these results to calculate
the other abundances.

\begin{table}[h]
\caption{Stages and equations} \label{resumen}
\begin{center}
\begin{tabular}{|c|c|c|}
\hline Stage& Equations &Final temperature \\ \hline

Until the weak interaction freeze-out&&\\ \hline

Until the production of \He becomes efficient&$
\dot{Y_n}=-2\dot{Y_{\alpha}}-Y_n[n] $&$2\dot{Y_{\alpha}}\sim
Y_n[n]$ \\ &$\dot{Y_d}=\dot{Y_3}=\dot{Y_T}=0$&  \\ \hline

Production of deuterium dominates the
&$\dot{Y_n}=-2\dot{Y_{\alpha}}$&$Y_n=Y_d$\\

rate of change of neutrons &$\dot{Y_d}=\dot{Y_3}=\dot{Y_T}=0$&
\\ \hline

Deuterium final abundance &$\dot{Y_d}=-2\dot{Y_{\alpha}}$&$T_9
\rightarrow 0$\\ &$\dot{Y_n}=\dot{Y_3}=\dot{Y_T}=0$&  \\ \hline

\end{tabular}
\end{center}
\end{table}

To calculate the final abundance of light elements it is necessary to
know the freeze-out temperature. The freeze-out of the production of
each element happens when the most important destruction reaction rate
equals to the expansion rate of the Universe. The dependence of the
freeze-out temperatures and final temperature of each stage with the
fundamental constants, will be calculated by deriving the equation that
determines each temperature.

Each section in this chapter will discuss the calculation of
abundances during a certain stage.


\subsection{Neutron abundance until the freeze-out of weak
  interaction, $T > 9.1 \times 10^9 \rm K$} 
\label{nuetrones}

For the calculation of neutron abundance we follow the analysis
performed by \citet{BBF88}. Let $\lambda_{pn}(T)$ be the
rate of weak process that convert protons into neutrons and
$\lambda_{np}(T)$ the rate of weak process that convert neutrons
into protons. The basic rate equation reads:
\begin{eqnarray}
\frac{dX}{dt}=\lambda_{pn}(t)(1-X(t))-\lambda_{np}(t)X(t)
\end{eqnarray}
where $t$ is the time, and $X$ is the ratio of the number of neutrons
to the total number of baryons. After changing variables $\left(
y={\frac{\Delta m}{T}}\right)$, the solution of the last equation can
be written as follows:
\begin{eqnarray}
X(y)=X_{eq}(y)+\int^y_0dy^{'} e^{y^{'}}\left[X_{eq}(y^{'})
\right]^2 e^{K(y)-K(y^{'})}
\end{eqnarray}
where
\begin{eqnarray}
K(y)&=&b\left[\frac{4}{y^3}+\frac{3}{y^2}+\frac{1}{y}+
\left(\frac{4}{y^3}+\frac{1}{y^2}\right)e^{-y}\right]; \hskip 1cm
b=255 \frac{M_{pl}}{\Delta m^2 \tau} \sqrt{\frac{45}{43\pi^3}}
\nonumber
\\ X_{eq}(y)&=&\frac{1}{1+e^y}
\end{eqnarray}
$\tau$ is the neutron mean life and $\Delta m=m_n-m_p$. In order to
obtain the asymptotic behavior, the limit $T\rightarrow 0$ or
$y\rightarrow \infty$ is taken:
\begin{eqnarray}
\label{xin} X(y=\infty)=\int^\infty_0dy^{'}
e^{y^{'}}X_{eq}\left(y^{'}\right)^2 e^{-K\left(y^{'}\right)}=0.151
\end{eqnarray}

In the last equation, only $b$ depends on the fundamental constants
through $\tau$ and $\Delta m$ (see sections \ref{masas} and
\ref{scattering} for the dependence of these quantities with the
fundamental constants). In such way, from equation \ref{xin}, we
obtain:
\begin{eqnarray}
\frac{\delta X(y=\infty)}{X(y=\infty)}&=
&-1.04 \frac{\delta G_F}{G_F}-2.361 \frac{\delta
\Lambda_{QCD}}{\Lambda_{QCD}}+1.386\frac{\delta \alpha}{\alpha}
\end{eqnarray}


\subsection{Until the production of $\He$ becomes efficient, $9.1
  \times 10^9 \rm K > T > 0.93 \times 10^9 \rm K  $} 
\label{seccion helio}

After the freeze-out of the weak interactions, the only change in the
neutron abundance is due to neutron decay. Therefore, the neutron
abundance in this stage reads:
\begin{eqnarray}
Y_n=X(y=\infty)\hskip 0.2cm e^{-t/\tau} =  \hskip 0.2cm e^{-0.198/T_9^2}
\end{eqnarray}

In the beginning of this stage there are no nucleus with two or more
nucleons, therefore it is a good approximation to consider: $Y_p \simeq
1-Y_n$. However, as the universe expands, the temperature goes down
and light nuclei formation begins. Therefore, at the end of this
stage, this expression is no longer valid.

In order to get a consistent solution of equation \ref{neutroncons}
\citep{Esma91}, it it is necessary to set all the rates equal to zero
with the exception of the largest rate which equals to $-2\dot
{Y_{\alpha}}-Y_n [n]$. In such way, the equations to solve in this
stage are:
\begin{eqnarray}
\label{yn1} \dot {Y_n}=-2\dot{Y_{\alpha}}-Y_n [n] \\
\dot{Y_d}=\dot {Y_3}=\dot {Y_T}=0
\end{eqnarray}
Table \ref{soluciones} shows the solutions.
\begin{table}
\caption{Solutions of the quasi static equilibrium equations for
each stage} \label{soluciones}
\begin{center}
\begin{tabular}{|c|c|c|}
\hline

$T_9$ & Nucleus & Solution \\ \hline

&\De&$Y_d=Y_n Y_p \frac{[npd\gamma]}{Y_{\gamma}[d\gamma np]} $\\
 \cline{2-3}

$9.1 > T_9 > 0.93 $ &\Tr&$Y_T=\frac{Y_d Y_p [pd3\gamma]+Y_d Y_d
[ddn3]+Y_T Y_p [pTn3]}{Y_d [d3p\alpha]+Y_n [n3pT]}$  \\
\cline{2-3}

&\Het& $Y_3=\frac{Y_n Y_3 [n3pT]+Y_d Y_d [ddpT]}{Y_d
[dTn\alpha]+Y_p [pTn3]}$\\ \hline

&\De& $Y_d= Y_n Y_p \frac{[npd\gamma]}{Y_{\gamma}[d\gamma np]}$ \\
\cline{2-3}

$0.93 > T_9 > 0.765$ & \Tr & $Y_3= Y_d \frac{[ddn3]}{[d3p\alpha]}$\\
\cline{2-3}

&\Het& $Y_T=  Y_d \frac{[ddpT]}{[dTn\alpha]}$\\ \hline

&D& $Y_d= Y_n Y_p \frac{[npd\gamma]}{Y_{\gamma}[d\gamma np]}$ \\
\cline{2-3}

$T_9 \rightarrow 0$&T& $Y_3= Y_d \frac{[ddn3]}{[d3p\alpha]}$\\
\cline{2-3}

&\Het& $Y_T=  Y_d \frac{[ddpT]}{[dTn\alpha]}$\\ \hline

\end{tabular}
\end{center}
\end{table}


When the production of $\He$ becomes efficient the stage ends. The
final temperature is given by is given by setting $\dot{Y_n}=0$ in
equation \ref{yn1}.  For this stage, we obtain $T_9^f=0.93$ and the
following results:
\begin{eqnarray}
Y_p^f&=&0.76 \hskip 2cm Y_d^f= 4.1 \times 10^{-4} \hskip 2cm
Y_T^f= 2.0 \times 10^{-5} \nonumber \\ Y_n^f&=&0.12 \hskip 2cm
Y_3^f= 5.8 \times 10^{-8} \hskip 2cm Y_{\alpha}^f=0.06 \nonumber
\end{eqnarray}
where $Y_i^f$ is the final abundance of each nucleus or nucleons on
this stage. It follows that the the abundances of $\De$, $\Tr$
and $\Het$ are negligible respect to the abundances of neutrons and
\He. This means:
\begin{eqnarray}
Y_p^f&=& 1-Y_n^f-Y_d^f-Y_T^f-2 Y_{\alpha}^f-2Y_3^f \nonumber \\
&\simeq&1-Y_n^f-2 Y_{\alpha}^f=1-2Y_n^f
\end{eqnarray}
Now, in order to calculate the dependence of the final temperature
with the fundamental constants for this stage, we derivate the
equation $2 \dot{Y_{\alpha}}=Y_n[n]$ with respect to the fundamental
constants and the temperature. In such way, we obtain:
\begin{eqnarray}
\frac{\delta T_9^f}{T_9^f}&=&0.068\frac{\delta \Omega_B h^2}{\Omega_B h^2}
-0.053 \frac{\delta G_F}{G_F}+0.063 \frac{\delta
\alpha}{\alpha}+0.871 \frac{\delta
\Lambda_{QCD}}{\Lambda_{QCD}}
\end{eqnarray}
where we also considered the dependence with the  baryon fraction. 
Finally, the dependence of the abundance of neutrons on
the fundamental constants and $\Omega_B h^2$ yields:
\begin{eqnarray}
\frac{\delta Y_n}{Y_n}=0.029\frac{\delta \Omega_B h^2}{\Omega_B h^2}
-1.522 \frac{\delta G_F}{G_F}+2.296 \frac{\delta
\alpha}{\alpha}-3.459 \frac{\delta \Lambda_{QCD}}{\Lambda_{QCD}}
\end{eqnarray}


\subsection{Final abundance of \He}

The next stage corresponds to the calculation until the rate of
production of deuterium dominates over the rate of change of
neutrons. However, the freeze-out temperature of $\He$ ($T=0.915
\times 10^9 \rm K $) is lower than the final temperature of the
previous stage but bigger than the final temperature of the next
one. Therefore, we calculate now the final abundance of \He. In this
case, the neutron number conservation equation reads:
\begin{eqnarray}
\label{principal} 2\dot{Y_{\alpha}} = Y_n [n]
\end{eqnarray}
For the others nucleus the quasi static equilibrium equation is valid
(see table \ref{soluciones}). The production of $\He$ is dominated by
$[dTn\alpha]$ and $[pT\gamma \alpha]$:
\begin{eqnarray}
\dot{Y_{\alpha}}=Y_d Y_T [dTn\alpha]+Y_p Y_T [pT\gamma
\alpha]=\left(Y_n Y_p \frac{[npd\gamma]}{Y_{\gamma}[d \gamma
np]}\right)^2 [ddpT]
\end{eqnarray}

After solving numerically for $T_9$ the equation (\ref{principal}), we
obtain $T_9^{\alpha} = 0.915$ and $Y_{\alpha}^f = 2 Y_n =0.238$.  When
the rate of $\He$ production equals to the rate of neutron
destruction, there is no more neutron that can form $\He$. Since this
happens earlier than the usual freeze-out-time, we use equation
\ref{principal} to calculate the freeze-out temperature. In such way,
the dependence of the freeze-out temperature on the fundamental
constants and $\Omega_B h^2$ yields:
\begin{eqnarray}
\frac{\delta T_9^{\alpha}}{T_9^{\alpha}}=0.061\frac{\delta \Omega_B h^2}{\Omega_B h^2}
-0.052\frac{\delta G_F}{G_F}+0.063 \frac{\delta
\alpha}{\alpha}+0.869 \frac{\delta \Lambda_{QCD}}{\Lambda_{QCD}}
\end{eqnarray}

Finally, since $Y_{\alpha}^c=2 Y_n$, we can express the
variation of the final abundance of $\He$ as a function of
fundamental constants and $\Omega_B h^2$:
\begin{eqnarray}
\frac{\delta Y_{\alpha}^c}{Y_{\alpha}^c}=0.029\frac{\delta \Omega_B h^2}{\Omega_B h^2}
-1.538 \frac{\delta G_F}{G_F}+2.324 \frac{\delta
\alpha}{\alpha}-3.496 \frac{\delta \Lambda_{QCD}}{\Lambda_{QCD}}
\end{eqnarray}


\subsection{Neutron cooking, $0.93 \times 10^9 \rm K > T > 0.765
  \times 10^9 \rm K$ } 

In this section we shall calculate the deuterium abundance as long as
the change of neutron dominates the deuterium production rate.  This
is valid until the production rate of deuterium dominates the rate of
change of neutrons, so this stage is over when $Y_n = Y_d$.  In this
stage, the neutron number conservation equation reads:
\begin{eqnarray}
\label{ynalpha} \dot {Y_n}=-2\dot{Y_{\alpha}}
\end{eqnarray}
For $\De$, $\Tr$ and $\Het$ we solve the quasi-static equilibrium
equations. The solutions are shown in table \ref{soluciones}. For
$\He$ we solve the complete equation but considering only the largest
production term $Y_d Y_T [dTn\alpha]$. Inserting all these solutions
in equation \ref{ynalpha}, we obtain:
%
%
%
%
\begin{eqnarray}
\dot{Y_n}=-2 \left(Y_n Y_p \frac{[npd\gamma]}{Y_{\gamma}[d\gamma
np]}\right)^2 [ddpT]
\label{neutequation}
\end{eqnarray}
where the initial condition is given by the final values of the
previous stage: $Y_n^0=0.12$ and $T_9^0=0.93$. We can write the
solution to the last equation as follows:
\begin{eqnarray}
\label{integral} Y_n=\left(\frac{1}{Y_n^0}+2\int^{t}_{t_{initial}}
\left(Y_p \frac{[npd\gamma]}{Y_{\gamma}[d\gamma np]}\right)^2
[ddpT] dt\right)^{-1}
\end{eqnarray}
After changing the integration variable to $T_9$ we perform the
integral numerically as a function of temperature. We also
compute the final temperature of this stage using the condition:
\begin{equation}
Y_n = Y_d
\label{secondstage}
\end{equation} 
We obtain:
\begin{eqnarray}
T_9^f=0.765 \hskip 2cm Y_n=6.4 \times 10^{-4}=Y_d
\end{eqnarray}
From \ref{secondstage} we obtain the dependence of the final
temperature of this stage with respect to the fundamental constants
and $\Omega_B h^2$:
\begin{eqnarray}
\frac{\delta T_9^f}{T_9^f}=0.031 \frac{\delta \Omega_B h^2}{\Omega_B h^2}+
0.015 \frac{\delta G_F}{G_F}-0.023 \frac{\delta
\alpha}{\alpha}+1.034 \frac{\delta \Lambda_{QCD}}{\Lambda_{QCD}}
\end{eqnarray}
Finally, the dependence of the final neutron and deuterium abundance
can be obtained from equation (\ref{integral}):
\begin{eqnarray}
\frac{\delta Y_d}{Y_d}=\frac{\delta Y_n}{Y_n}=-1.099 \frac{\delta \Omega_B h^2}{\Omega_B h^2}
-0.058 \frac{\delta G_F}{G_F}+1.871 \frac{\delta
\alpha}{\alpha}-0.488 \frac{\delta \Lambda_{QCD}}{\Lambda_{QCD}}
\end{eqnarray}



\subsection{Deuterium cooking, $T \rightarrow 0$}

For temperatures lower than $T_9 = 0.765$, the largest production rate
corresponds to deuterium. Therefore, we set all other derivatives to
zero in equation \ref{neutroncons}. Since the largest term for
deuterium destruction is tritium production, the equation to solve is:
\begin{eqnarray}
\dot {Y_d}=-2 Y_d Y_d [ddpT]
\end{eqnarray}
with the initial condition $Y_d^0=6.4 \times 10^{-4}$ on
$T_9^0=0.765$.  Since this equation has the same form of equation
\ref{neutequation}, the solution reads:
\begin{eqnarray}
\label{ydfinal}
 Y_d=\left(\frac{1}{Y_d^0}+2\int^{t}_{t_{initial}}
[ddpT] dt\right)^{-1}
\end{eqnarray}

In order to calculate the deuterium final abundance we consider the
limit $T \rightarrow 0$ ( $t \rightarrow \infty$). We obtain
numerically, the deuterium final abundance $Y_d^f=2.410 \times
10^{-5}$. On the other hand, the dependence of the deuterium final
abundance with the fundamental constants and $\Omega_B h^2$ can be
calculated by deriving equation \ref{ydfinal}:
%
%
\begin{eqnarray}
\label{variacion de yd final}
\frac{\delta Y_d^c}{Y_d^c}=-1.072\frac{\delta \Omega_B h^2}{\Omega_B h^2}
-0.036 \frac{\delta G_F}{G_F}+2.320 \frac{\delta
\alpha}{\alpha}+0.596 \frac{\delta \Lambda_{QCD}}{\Lambda_{QCD}}
\end{eqnarray}


\subsection{Final abundances}

Here we calculate the freeze-out temperature and final abundances of
$\Het$, $\Tr$, $\Lis$, $\Be$ and $\Li$ and the dependence of these
quantities with the fundamental constants. In order to calculate any
light element abundance it is necessary to solve the quasi-static
equilibrium equation:
\begin{eqnarray}
\dot{Y_i}=0
\end{eqnarray}
We solve these equations considering only the most relevant
reactions. In table \ref{importancia} we show the quasi-static
equilibrium solutions.

\begin{table}[h]
\caption{Quasi-static equilibrium solutions} \label{importancia}
\begin{center}
\begin{tabular}{|c|c|}
\hline

Nucleus &  Quasi-static equilibrium solutions
\\ \hline

\Het &$Y_3=\frac{ Y_d [ddn3]}{[d3p\alpha]}$ \\\hline

\Tr &$Y_T=\frac{ Y_d [ddpT]}{[dTn\alpha]}$ \\\hline

\Lis &$Y_6= \frac{Y_d Y_{\alpha} [d \alpha 6 \gamma]}{Y_p
[p6T\alpha]}$
\\\hline

$\Be$ &$Y_B=\frac{Y_3 Y_{\alpha} [3 \alpha B \gamma]}{Y_n
[nBp7]}$\\\hline

\Li &$Y_7=\frac{Y_n Y_B [nBp7]+Y_T Y_{\alpha} [T \alpha 7
\gamma]}{Y_p [p7 \alpha \alpha]}$\\\hline
\end{tabular}
\end{center}
\end{table}

In order to compute the freeze-out temperature, we set the largest
rate of destruction $\Gamma$ of each equation that governs the
abundance of the light elements equal to the universe expansion rate
$H$:
\begin{eqnarray}
\label{h} \Gamma = H =\frac{1}{356} T_9^2 seg^{-1}
\end{eqnarray}
Table \ref{temperatura} shows the different freeze-out temperatures
and their dependence on fundamental constants which is calculated
deriving the previous equation. Using the freeze-out temperature we
calculate the final abundance of the different nucleus and their
dependence on the fundamental constants and $\Omega_B h^2$. In table
\ref{varia} we show these results.

\begin{table}[h]
\caption{Freeze-Out temperature and their dependence on fundamental
constants, $\frac{\delta T_9^i}{T_9^i}=W\frac{\delta
G_F}{G_F}+R\frac{\delta \alpha} {\alpha}+T \frac{\delta
\Lambda_{QCD}}{\Lambda_{QCD}}+J \frac{\delta \Omega_B
h^2}{\Omega_B h^2}$} \label{temperatura}
\begin{center}
\begin{tabular}{|c|c|c|c|c|c|c|}
\hline

Nucleus&Equation& $T_9^{Freeze-Out}$&$W$&$R$&$T$&$J$\\ \hline

$\Het$ &$Y_d[d3p\alpha]=H$&
$0.403$&$0.008$&$-0.510$&$1.168$&$0.016$\\ \hline

$\Tr$ &$Y_d [dTn\alpha]=H$& $0.105$&$0.009$&$0.122
$&$1.181$&$0.018$\\ \hline

$\Lis $& $Y_p [p63\alpha]=H$&$0.069$&$-0.076
$&$1.962$&$1.118$&$-0.156$\\ \hline

$\Be$&$Y_n[nBp7]=H$&$0.319$&$0.217$&$-0.712$&$1.39$&$0.350$\\
\hline

$\Li $&$Y_p [p7 \alpha
\alpha]=H$&$0.185$&$-0.088$&$1.692$&$0.946$&$-0.182$\\ \hline
\end{tabular}
\end{center}
\end{table}

\begin{table}[h]
\caption{Abundances and their dependence on fundamental constants,
$\frac{\delta Y_i^f}{Y_i^f}=A\frac{\delta G_F}{G_F}+B\frac{\delta
\alpha} {\alpha}+C \frac{\delta
\Lambda_{QCD}}{\Lambda_{QCD}}+D\frac{\delta \Omega_B h^2}{\Omega_B
h^2}$} \label{varia}
\begin{center}
\begin{tabular}{|c|c|c|c|c|c|}
\hline $Y_i^f$&Abundance&$A$&$B$&$ C$&$D$\\ \hline

$\D $&$2.741 \times 10^{-5}$&$-0.036 $&$2.320 $&$0.596$
&$-1.072$\\ \hline

$^3He$&$6.95 \times10^{-6}$&$-0.051 $&$0.983 $&$0.999$&$-1.102$
\\ \hline

$\T$ &$1.21\times 10^{-7}$& $-0.041 $&$0.252 $&$0.941 $
&$-1.083$\\ \hline

$^4He$&$0.238$&$-1.538 $&$2.323 $&$-3.497$ &$0.029$\\ \hline

$^6Li $&$5.7 \times 10^{-14}$ &$-2.061 $&$7.414
$&$-3.462$&$-1.047$
\\ \hline

$^7Be$&$5.60 \times 10^{-10}$ &$ -0.172 $&$-9.450 $&$-1.038$
&$2.209$
\\ \hline

$^7Li $&$2.36\times 10^{-10}$&$-0.720 $&$1.824 $&$-3.411
$&$0.068$\\ \hline
\end{tabular}
\end{center}
\end{table}

\section{Results and  discussion}
\label{analisis}

In this section we compare the theoretical predictions of the
abundances of the light elements obtained in the last section with
observational data.

In section \ref{abundancias} we have obtained $7$ equations of the form:
\begin{eqnarray}
\label{ecuaciones} \frac{\delta Y_i^f}{Y_i^f}=A_i\frac{\delta
G_F}{G_F}+B_i\frac{\delta \alpha} {\alpha}+C_i \frac{\delta
\Lambda_{QCD}}{\Lambda_{QCD}}+D_i \frac{\delta \Omega_B
h^2}{\Omega_B h^2}
\end{eqnarray}
where $A_i$, $B_i$, $C_i$ and $D_i$ are constant coefficients (see
table \ref{varia}), $\frac{\delta
Y_i}{Y_i}=\frac{Y_i^{obs}-Y_i^{SBBN}}{Y_i^{SBBN}}$; and $Y_i^{SBBN}$
and $Y_i^{obs}$ are the theoretical and observed abundance
respectively.

However, independent observational data are only available for the
abundances of $\De$, $\Het$, $\He$ and $\Li$. In table \ref{obs} we
show the independent data we consider in this work.  For a recent review of all observational available data on primordial abundances see \citet {RO04}. On the other
hand, recent papers \citep{coc04,cyburt04} have brought the attention
to the errors introduced by the values of the cross sections involved
in the calculation of the abundances. \citet{cyburt04} has also
analyzed the propagation through the theoretical abundances, yielding
a ``theoretical'' percent error of 5\%. In the original work of
\citet{Esma91}, the error introduced by the semi-analytical method is
estimated to be of order $5 \%$. Therefore, we will add in order to
solve system \ref{ecuaciones}, to the errors of table \ref{obs} an
error of order $10 \%$.

First we perform a test to check the consistency of the data
\citep{RV86}.  For each group of data ($Y_i$) belonging to the same
abundance, we calculate the weighted averaged value $\overline{Y}$ and
its corresponding error $\sigma_i$. Then we compute:
\begin{equation}
\chi^2 = \sum_i \frac{(Y_i - \overline{Y})^2}{\sigma_i^2}
\end{equation}
If the errors are Gaussian distributed, the expected value of $\chi^2$
is ($k-1$) where $k$ is the number of data in each group.
Furthermore, the corresponding ideogram of each group of data
\citep{data02}, should be a Gaussian.  It follows from figure
\ref{ideogramas} and from the calculation of $\chi^2$ that $\De$ and
$\He$ data are not Gaussian distributed. However, since $\Theta=
\sqrt{\frac{\chi^2}{k-1}}$ is not that greater than one, we can use
the data but increasing the observational error by a factor
$\Theta$. The values of $\Theta$ are $2.4$ for $\De$, $2.33$ for
$\He$.

We assume that any difference between the theoretical abundance and
the observational abundance is due to the variation of fundamental
constants. In such way, the solution of system (\ref{ecuaciones})
gives a constraint to this variation. The solution is given by
\citep{Arley}:
\begin{eqnarray}
\label{variaciones} \frac{\delta
\alpha_i}{\alpha_i}=\left[\left(B^t P B\right)^{-1}B^t P\delta
\right]_{i} \pm \sqrt{\left[\left(B^t P
B\right)^{-1}\right]_{ii}}s
\end{eqnarray}
where $B$ is the $n\times 4$ matrix, $n$ is the number of
observational data:
\begin{eqnarray}
B=\left (
\begin{array}{cccc}
A_{1} & B_{1} & C_{1} & D_{1}\\ A_{2} & B_{2} & C_{2}& D_{2}\\
\vdots & \vdots & \vdots& \vdots \\ A_{n} & B_{n} & C_{n}& D_{n}
\end{array}
\right)
\end{eqnarray}
$\delta$ is the $n\times 1$ matrix:
\begin{eqnarray}
\delta=\left (
\begin{array}{c}
\frac{\delta Y_1}{Y_1}\\ \frac{\delta Y_2}{Y_2}\\ \vdots \\
\frac{\delta Y_n}{Y_n}
\end{array}
\right)
\end{eqnarray}
and $P$ is the $n \times n$ matrix of weight:
\begin{eqnarray}
P=\left (
\begin{array}{cccc}
p_1 & 0&\ldots&0\\ 0&p_2&\ldots&0\\ \vdots &\vdots & \ddots
&\vdots\\ 0&0&\ldots&p_n
\end{array}
\right)
\end{eqnarray}
where $p_i=\frac{1}{\sigma_i^2}$ and $\sigma_i$ are the observational errors.


The most accurate estimation of $\Omega_B h^2$
arrives from constraining parameters with data from the CMB provided
by WMAP \citep{wmapest}. Fixing the baryon fraction with the WMAP value (i.e. setting $D_i=0$), the results of solving the system \ref{ecuaciones} with all data
listed in table \ref{obs} are shown in table \ref{resultadoswmap}. These results are
consistent within $3 \sigma$ with variation of the fundamental
constants. On the other
hand, the results considering only variation of the fine structure
constant are shown in table \ref{alphawmap}. These results are
consistent with no variation of $\alpha$ within $3 \sigma$.
 In order
to rule out any systematic error of the data, we computed the solution
of system \ref{ecuaciones} again but excluding one group of data at
each time. Again, the results are consistent with variation of the
fundamental constants in all cases but in the case where the $\Li$
data were excluded (see tables \ref{resultadoswmap} and \ref{alphawmap}).

Even though, the WMAP estimate of the baryon density is the most accurate one, it is still affected by degeneracies with other cosmological parameters \citep{wmapest}.  Therefore, we 
added  an independent estimation of $\Omega_B h^2$
in our analysis. In appendix \ref{rayos} we
use data from X-ray measurements,
galaxy surveys and cepheids calibration in order to get an independent value of the baryon density. Furthermore, we computed again the results of sections \ref{scattering} and \ref{abundancias}, changing the value of $\Omega_B h^2$ to $0.0223$. This value is the weighed mean value between the WMAP estimate and the value of appendix \ref{rayos}. However, we found no difference in the value of the coefficients of the variation of fundamental constants and $\Omega_B h^2$.   
The results obtained solving  system \ref{ecuaciones}  including
both estimates for the baryon fraction (i. e.  $D_i \ne 0$ ) show no significant difference with respect to  the case where only the WMAP value was considered (see tables {\ref{resultados} and \ref{resulalpha}). Furthermore, in order to check for consistency of our method, we solved again system \ref{ecuaciones} allowing only for variation of $\Omega_B h^2$ with respect to the weighed mean value (i.e. $A_i=B_i=C_i=0$). These results are shown in table \ref{resulob}. On the other hand, in order to learn about the degeneracies of the fundamental constants within the BBN model, we computed  the correlation coefficients from the error matrix. We find that there is high correlation between $\alpha$ and $\Lambda_{QCD}$, $\alpha$ and $G_F$ and $\Lambda_{QCD}$ and $G_F$, while the correlation between other pairs of parameters is not significant. 

In order to understand the discrepancy of the results obtained with and without the $\Li$ data, we computed
the relative residuals \citep{Arley}, and their respective theoretical
and empirical probability in both cases.  Figure \ref{KS} shows that in the case
where both the variation of the fundamental constants and the
deviation of $\Omega_B h^2$ from the WMAP estimate is considered, the
theoretical and empirical probability distributions are very similar,
while in the case where only the deviation of $\Omega_B h^2$ is
considered, there is slight difference between the empirical
probabilities (both with all data and excluding $\Li$ data) and the
theoretical probability. Including the variation of fundamental
constants gives more degrees of freedom to system
\ref{ecuaciones}. Therefore, we suspect that the possible non reported
systematic uncertainties ``hide'' under the variation of the
fundamental constants.  On the other hand, we performed a
Kolmogorov-Smirnov (K-S) test, in order to check the goodness of our
fit. For the results obtained considering variation of all constants and $\Omega_B h^2$,
we obtain a probability of $21\%$ to obtain a worse fit, while
excluding the $\Li$ data the probability lowers to $11\%$. On the
other hand, if we only consider the deviation of $\Omega_B h^2$ with
respect to the WMAP data, we obtain a
probability of $99\%$ for all data, while excluding the $\Li$ data
gives a $49 \%$ of probability to get a worse fit. However, we
consider the results of the K-S test only indicative, since even
though the data considered are independent the residuals are not.

We mentioned in the introduction the disagreement between the $\Li$
observational abundances with the $\De$ observational abundance and
WMAP estimate of the baryon density. \citet{richard05} claim that a
better understanding of turbulent transport in the stars is necessary
to understand this discrepancy. Moreover, \citet{MR04} have reanalyzed
the $\Li$ data with an improved infrared flux method temperature
scale, obtaining values that are marginally consistent with the WMAP
estimate. However, solving system \ref{ecuaciones} with the $\Li$
abundance taken from their work, does not change in a significant way
our results.

We adopt the conservative criterion that the third and fourth column
of tables \ref{resultadoswmap} and  \ref{resultados} are the constraints on the variation of the
constants we obtain with the method and hypothesis described in this
paper. We also consider that more observations of $\Li$ are needed in
order to arrive to stronger conclusions.  However, if the all data are
correct, this analysis shows that varying coupling constants may solve
the concordance problem between BBN and CMB.  Our results within 2
$\sigma$ are consistent with the analysis performed by \citet{ichi04},
where only the variation of $\alpha$ and a non standard expansion
rate.

\begin{table}[h]
\caption{Theoretical abundances in the standard model the WMAP estimate  $\Omega_B h^2=0.0224$} 
\label{SBBN}
\begin{center}
\begin{tabular}{|c|c|}
\hline Nucleus & $Y_i^{SBBN} \pm \delta Y_i^{SBBN}$  \\ \hline
$\D$&$\left(2.51\pm 0.37\right)\times 10^{-5}$\\ \hline
$^3He$&$\left(1.05 \pm 0.15\right)\times 10^{-5}$\\ \hline
$^4He$&$0.2483\pm 0.0012$\\\hline 
$^7Li$&$\left(5.0 \pm 0.3\right)\times 10^{-10}$\\ \hline
\end{tabular}
\end{center}
\end{table}

\begin{table}[h]
\caption{Observational abundances used in this work} \label{obs}
\begin{center}
\begin{tabular}{|c|c|c|}
\hline

Nucleus & $Y_i^{obs} \pm \delta Y_i^{obs}$ & Cite  \\ \hline

D& $\left (1.65 \pm 0.35\right ) \times 10^{-5}$&\cite{pettini}\\
\hline

D&$\left (2.54 \pm 0.23\right ) \times 10^{-5}$&\cite{omeara}
\\ \hline

D&$\left (2.42^{+0.35}_{-0.25}\right ) \times
10^{-5}$&\cite{kirkman} \\ \hline

D&$\left (3.25 \pm 0.3\right ) \times 10^{-5}$& \cite{burles2}\\
\hline

D&$\left (3.98^{+0.59}_{-0.67}\right ) \times 10^{-5}$&
\cite{burles1}\\ \hline

D&$\left (1.6^{+0.25}_{-0.30}\right ) \times 10^{-5}$&
\cite{Crighton04}\\ \hline

$^3\rm{He}$&$\left(1.1 \pm 0.2\right) \times 10^{-5}$&
\cite{Bania}
\\ \hline

$^4\rm{He}$& $0.244 \pm 0.002$&\cite{izotov1} \\ \hline

$^4\rm{He}$&$0.243 \pm 0.003$&\cite{izotov2}\\ \hline

$^4\rm{He}$&$0.2345 \pm 0.0026$ &\cite{peimbert1} \\ \hline

$^4\rm{He}$&$0.232 \pm 0.003$&\cite{oli95}\\ \hline

$^7\rm{Li}$&$\left(1.23^{+0.68}_{-0.32}\right) \times
10^{-10}$&\cite{ryan} \\ \hline

$^7\rm{Li}$&$\left(1.58^{+0.24}_{-0.20}\right) \times 10^{-10}$
&\cite{bonifacio1}\\ \hline

$^7\rm{Li}$&$\left(1.73 \pm 0.05\right) \times 10^{-10}$
&\cite{bonifacio2}\\ \hline

$^7\rm{Li}$&$\left(2.19^{+0.30}_{-0.26}\right) \times 10^{-10}$
&\cite{bonifacio3}\\ \hline
\end{tabular}
\end{center}
\end{table}

\begin{figure}
\begin{center}
\includegraphics[scale=0.4,angle=-90]{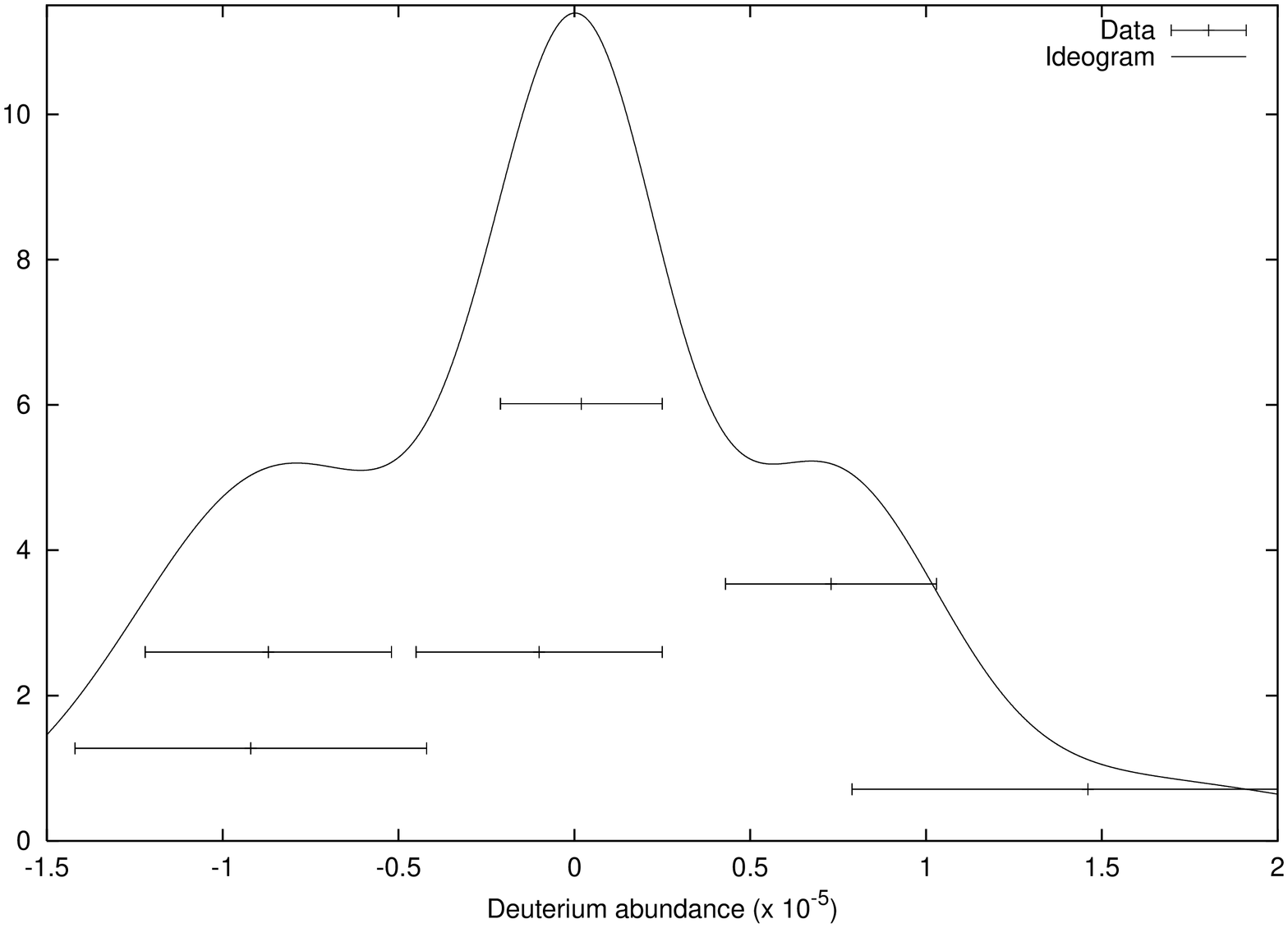}
\hspace{1cm}
\includegraphics[scale=0.4,angle=-90]{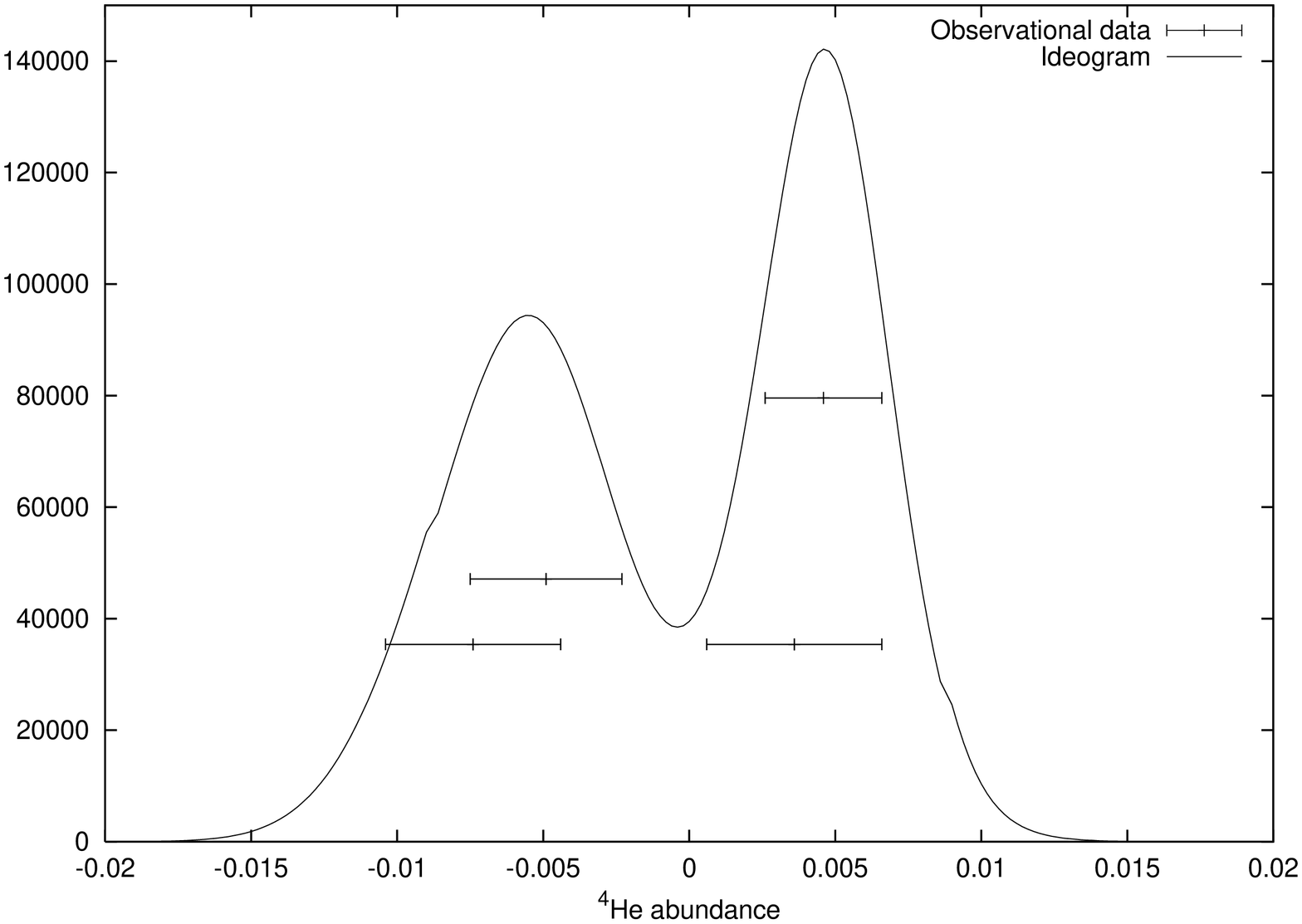}
\end{center}
\caption{Ideograms for $\De$ and $\He$.}
\label{ideogramas}
\end{figure}

\begin{figure}
\begin{center}
\includegraphics[scale=0.6]{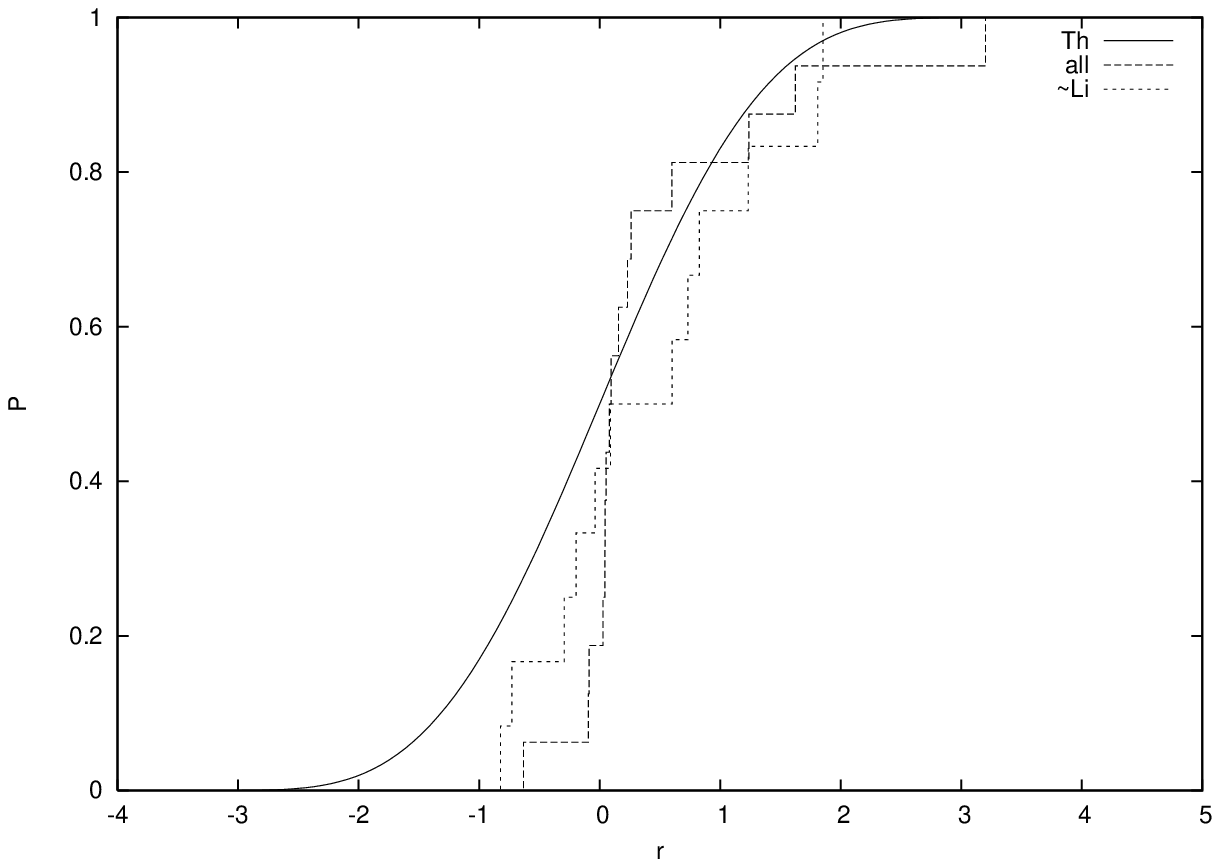}
\hspace{1cm}
\includegraphics[scale=0.6]{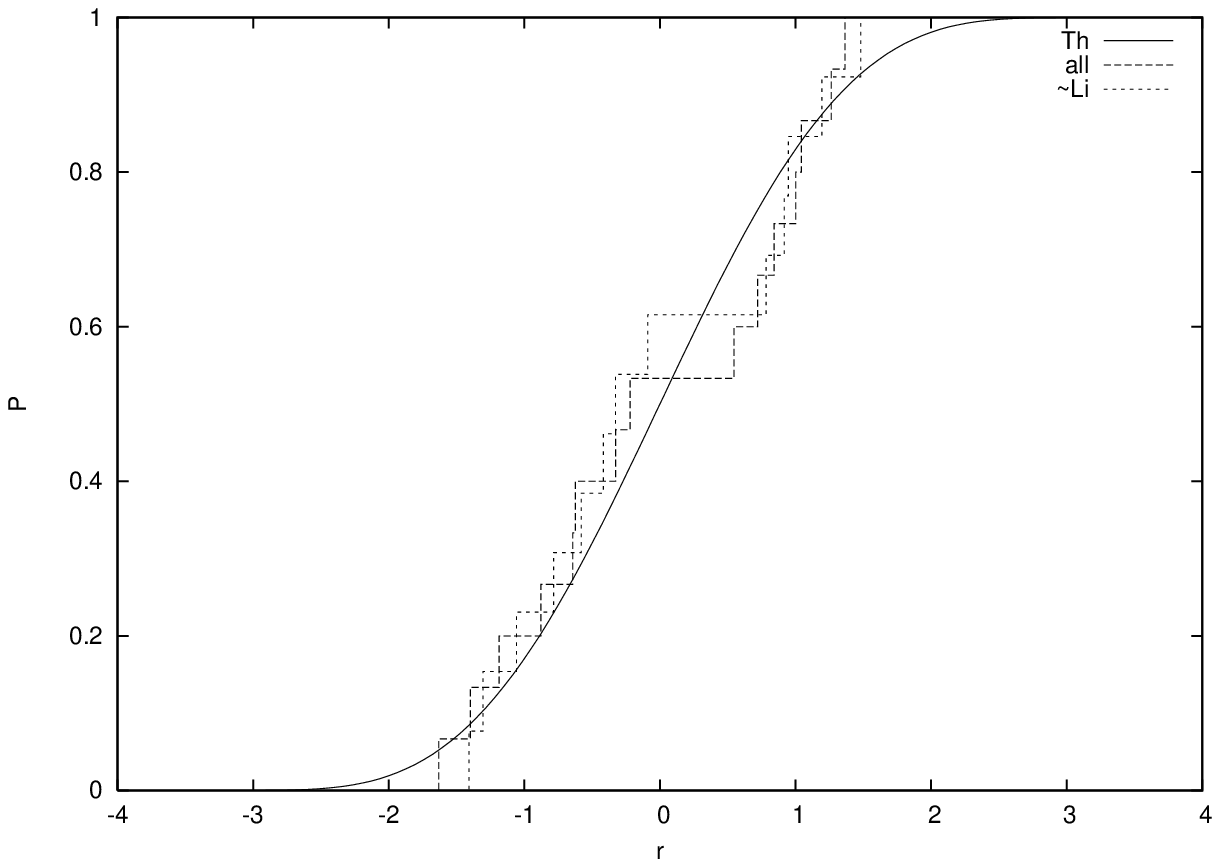}
\end{center}
\caption{The full line shows the theoretical probability of the
residuals, the dotted line shows the empirical probability computed with all
data and the dotted line shows the empirical probability computed with
all data but $\Li$. Left: Only deviation of $\Omega_B h^2$ with
respect to its mean value is considered; Right: variation of all
constants and deviation of $\Omega_B h^2$ from the mean value is
considered.}
\label{KS}
\end{figure}

\begin{table}[h]
\caption{Constraints on the variation of fundamental constants ($\Omega_B h^2 = 0.0224$). }
\label{resultadoswmap}
\begin{center}
\begin{tabular}{|c|c|c|c|c|}
\hline& \multicolumn{2}{c}{All data} \vline &
\multicolumn{2}{c}{All data but $\Li$} \vline\\ \hline

&Value& $\sigma$&Value& $\sigma$ \\ \hline

$\frac{\delta G_F}{G_F}$&-0.886&0.053&-0.257&0.659
\\\hline

$\frac{\delta \alpha}{\alpha}$&-0.136&0.041&-0.054&0.097
\\\hline

$\frac{\delta  \Lambda_{QCD}}{
\Lambda_{QCD}}$&0.309&0.023&0.087&0.233
\\ \hline
\end{tabular}
\end{center}
\end{table}

\begin{table}[!h]
\caption {Constraints on  on the variation of  $\alpha$ ($\Omega_B h^2 = 0.0224$).}
\label{alphawmap}
\begin{center}
\begin{tabular}{|c|c|c|c|c|}
\hline& \multicolumn{2}{c}{All data} \vline &
\multicolumn{2}{c}{All data but $\Li$} \vline\\ \hline &Valor&
$\sigma$&Value& $\sigma$
\\ \hline

$\frac{\delta \alpha}{\alpha}$ &-0.041&0.024&-0.015&0.005
\\\hline
\end{tabular}
\end{center}
\end{table}

\begin{table}[h]
\caption{Constraints on the variation of the fundamental constants using two independent estimates for the baryon fraction.}
\label{resultados}
\begin{center}
\begin{tabular}{|c|c|c|c|c|}
\hline

&\multicolumn{2}{c}{All data}\vline &\multicolumn{2}{c}{All data
but $\Li$ }\vline \\ \cline{2-5}

& Value& $\sigma$&Value& $\sigma$ \\ \hline

$\frac{\delta \Omega_B h^2}{ \Omega_B h^2}$&0.004 &0.036 & 0.0005
&0.039\\ \hline

$\frac{\delta G_F}{G_F}$& -0.886&0.050&-0.258&0.64\\
\hline

$\frac{\delta \alpha}{\alpha}$& -0.134 &0.044 &-0.053&0.095\\
\hline

$\frac{\delta  \Lambda_{QCD}}{ \Lambda_{QCD}}$& 0.310 &0.023
&0.087&0.229\\ \hline

\end{tabular}
\end{center}
\end{table}

\begin{table}[h]
\caption{Constraints on the variation of  $\alpha$ using two independent estimates for the baryon fraction.}
\label{resulalpha}
\begin{center}
\begin{tabular}{|c|c|c|c|c|}
\hline

&\multicolumn{2}{c}{All data}\vline &\multicolumn{2}{c}{All data
but $\Li$ }\vline \\ \cline{2-5}

& Value& $\sigma$&Value& $\sigma$ \\ \hline

$\frac{\delta \alpha}{\alpha}$& -0.086&0.034&-0.015&0.005 \\
\hline

\end{tabular}
\end{center}
\end{table}

\begin{table}[h]
\caption{Constraints on the deviations of  $\Omega_B h^2$ respect
to
  the mean value considered in this work (0.0223).}
\label{resulob}
\begin{center}
\begin{tabular}{|c|c|c|c|c|}
\hline

&\multicolumn{2}{c}{All data}\vline &\multicolumn{2}{c}{All data
but $\Li$ }\vline \\ \cline{2-5}

& Value& $\sigma$&Value& $\sigma$ \\ \hline

$\frac{\delta \Omega_B h^2}{\Omega_B h^2}$
&-0.085&0.294&-0.014&0.054

\\ \hline
\end{tabular}
\end{center}
\end{table}

\appendix
\section{Appendix I}
\label{rayos}

In this appendix, we combine independent astronomical data in order to
obtain and independent estimation of the baryon density. From
measurements of hot gas in clusters it possible to obtain an estimate
of $\frac{\Omega_B}{\Omega_m} h^{3/2}$. 
 
\citet{Ettori03} has brought the attention to the fact that the
contribution from baryons in galaxies and ``exotic sources'' like
intergalactic stars and baryonic dark matter are not considered in the
results obtained from measurements of hot gas in
clusters. Furthermore, \citet{chandra03} have estimated the
contribution from the galaxies as follows: $f_{gal}= 0.15 h^{3/2}
f_{gas}$ while the ``exotic'' contribution has been estimated in
$f_{exotic} = 0.3 f_{gal}$ \citep{Ettori03}. Therefore, we add to the
estimation of the baryon fraction done by \citet{chandra03} the
contribution from galaxies, yielding the following value:

\begin{equation}
\frac{\Omega_B}{\Omega_m} h^{3/2} = 0.0737 \pm 0.0143 
\end{equation} 
The values of the other estimates \citep{xmm02,rosat03} are contained
within the error in this estimation.

\begin{table}
\caption{Observational data used to perform an estimate of the baryon density}
\label{tablabarion}
\begin{center}
\begin{tabular}{|c|c|}
\hline
   $\frac{\Omega_B}{\Omega_m} h^{3/2}$ & Cite \\
\hline
 $0.067 \pm 0.03$ & \citep{chandra03} \\
 $ 0.073 \pm 0.013$ & \citep{xmm02}  \\
 $0.056 \pm 0.007$ &\citep{rosat03} \\
\hline
 $\Omega_m h$ & Cite \\
\hline
$0.20\pm 0.03$ & \citep{2dF01} \\
 $0.207 \pm 0.030$ & \citep{Sloan04}  \\
\hline
 h & Cite \\
\hline
$ 0.72 \pm 0.08$ & \citep{hst01}\\
\hline
\end{tabular}
\end{center}
\end{table}

On the other hand, $\Omega_m h$ has been estimated from large redshift galaxy surveys like Sloan Digital Sky Survey \citep{Sloan04} and 2dF Galaxy Redshift Survey \citep{2dF01}, while the most stringent bound on the Hubble constant follows from cepheid calibration  \citep{hst01}.
Thus, combining all these data (see table \ref{tablabarion}) , and after propagating errors, we obtain the
following value for the baryon density:
\begin{equation}
\Omega_B h^2 = 0.017 \pm 0.007
\end{equation}
 
This value is less accurate that the estimation done with the data of
WMAP \citep{wmapest}
\begin{equation}
\Omega_B h^2 = 0.0224 \pm 0.0009
\end{equation}
but we will consider it in order to have an independent data of this quantity.





\newpage

\bibliography{bibliografia}
\bibliographystyle{astron}

\end{document}